\documentclass[journal,transmag]{IEEEtran}
\IEEEoverridecommandlockouts
\usepackage{amsmath}
\usepackage{amssymb}
\usepackage{mathrsfs}
\usepackage{amsfonts}
\usepackage[linesnumbered, ruled]{algorithm2e}
\usepackage{setspace}
\usepackage{color}
\usepackage{cite}
\usepackage{multirow}
\usepackage{tabularx}
\usepackage{amsthm}
\usepackage{algorithmic}
\usepackage{graphicx}
\usepackage{multicol}
\usepackage{subfigure}
\usepackage{bm}
\usepackage{booktabs}

\begin{document}
%{\setlength\abovedisplayskip{1pt}
%\setlength\belowdisplayskip{1pt}}

\setlength{\textfloatsep}{0\baselineskip plus 0.2\baselineskip minus 0.2\baselineskip}

\title{QoS-aware User Grouping Strategy for Downlink Multi-Cell NOMA Systems}

%\author{\IEEEauthorblockN{Fengqian Guo, Hancheng Lu}
%   \IEEEauthorblockA{
%       The signal Network Lab of EEIS Department, USTC, Hefei, China, 230027 \\
%		Email: fqguo@mail.ustc.edu.cn, hclu@ustc.edu.cn%, darenzhu@mail.ustc.edu.cn, hwu2014@mail.ustc.edu.cn
%	}
%}

\author{Fengqian~Guo,~%~\IEEEmembership{Member,~IEEE,}
        Hancheng~Lu,~%~\IEEEmembership{Fellow,~OSA,}
        Xiaoda~Jiang,~%~\IEEEmembership{Life~Fellow,~IEEE}
        Ming~Zhang,~%,~\IEEEmembership{Life~Fellow,~IEEE}% <-this % stops a space
        Jun~Wu,
        and~Chang Wen Chen
\IEEEcompsocitemizethanks{\IEEEcompsocthanksitem
This work was supported by National Key R\&D Program of China under Grant 2020YFA0711400 and National Science Foundation of China under Grant  61631017, 61771445, 91538203.

Fengqian Guo, Hancheng Lu, Xiaoda Jiang, and Ming Zhang are with the Key Laboratory of Wireless-Optical Communications, Chinese Academy of Sciences, University of Science and Technology of China, Hefei 230027, China (e-mail: fqguo@mail.ustc.edu.cn; hclu@ustc.edu.cn; jxd95123@mail.ustc.edu.cn; mzhang95@mail.ustc.edu.cn).

Jun Wu is with School of Computer Science, Fudan University, Shanghai 200433, China (email: wujun@fudan.edu.cn).

%Chang Wen Chen is with the Department of Computing, Hong Kong Polytechnic University, Hung Hom, Kowloon, Hong Kong SAR, China (e-mail: changwen.chen@polyu.edu.hk).\protect\\

Chang Wen Chen is with the Department of Computing, Hong Kong Polytechnic University, Hung Hom, Kowloon, Hong Kong SAR, China (e-mail: chencw@buffalo.edu).\protect\\
}% <-this % stops an unwanted space
}

\maketitle

%\IEEEtitleabstractindextext{%
\begin{abstract}
In multi-cell non-orthogonal multiple access (NOMA) systems, designing an appropriate user grouping strategy is an open problem due to diverse quality of service (QoS) requirements and inter-cell interference. In this paper, we exploit both game theory and graph theory to study QoS-aware user grouping strategies, aiming at minimizing power consumption in downlink multi-cell NOMA systems. Under different QoS requirements, we derive the optimal successive interference cancellation (SIC) decoding order with inter-cell interference, which is different from existing SIC decoding order of increasing channel gains, and obtain the corresponding power allocation strategy. Based on this, the exact potential game model of the user grouping strategies adopted by multiple cells is formulated. We prove that, in this game, the problem for each player to find a grouping strategy can be converted into the problem of searching for specific negative loops in the graph composed of users. Bellman-Ford algorithm is expanded to find these negative loops. Furthermore, we design a greedy based suboptimal strategy to approach the optimal solution with polynomial time. Extensive simulations confirm the effectiveness of grouping users with consideration of QoS and inter-cell interference, and show that the proposed strategies can considerably reduce total power consumption comparing with reference strategies.

\end{abstract}

%\newpage
%}

\IEEEdisplaynontitleabstractindextext

% \fi
\IEEEpeerreviewmaketitle

\section{Introduction}
\newtheorem{def1}{\bf Definition}
\newtheorem{thm1}{\bf Theorem}
\newtheorem{lem1}{\bf Lemma}
\newtheorem{cor1}{\bf Corollary}

%\IEEEPARstart{A}{s}
As the mobile communications develop with great speed, higher system throughput and massive connectivity are needed for 5th-generation (5G) mobile communication systems \cite{vaezi2019multiple, Guiyongqiang2020MC,fu2020secure}. To address these challenges, non-orthogonal multiple access (NOMA) is recognized as a potential technique \cite{vaezi2019multiple}. In NOMA systems, signals of multi-users can be transmitted on the same frequency domain simultaneously by superposition coding(SC)\cite{8790780}. {These signals are distinguished according to their different transmit power.} Then successive interference cancellation (SIC) is used to detect and decode these signals \cite{Shin2017}.

%\textcolor{red/blue/green/black/white/cyan/magenta/yellow}{text}
{Due to the power domain multiplexing, NOMA can serve more users than traditional orthogonal multiple access(OMA) and provide higher multi-user diversity gain and spectrum efficiency\cite{Choi2017,ZhangMing2021MC}. There have been many attempts on power allocation to optimize the performance of NOMA, including energy efficiency (EE)\cite{zhang2016energy, Moltafet2018}, fairness\cite{Cui2016, zhu2017optimal} and sum rate\cite{Al-Abbasi2017, Xing2018}.
Power allocation in multi-cell NOMA networks has been studied in \cite{Yang2018,Fu2017}. However, there will be severe inter-user interference and error propagation of SIC if too many users sharing the same time-frequency resource. In addition, more users sharing the same time-frequency resource means higher computation complexity of SIC at the receiver. User grouping has been considered as an effective way to alleviate inter-user interference and error propagation, achieving a balance between the system performance and computation complexity \cite{vaezi2019multiple}. With user grouping, users are divided in groups, and the radio resources among different groups are orthogonal. Inter-user interference and power consumption are closely related to user grouping, thus total transmit power can be reduced by an appropriate user grouping strategy\cite{dzg1, Cui2018}.}

Many studies have been done on user grouping in NOMA systems \cite{dzg1, Lv2018, Ali1, Kiani2018, Zhou2018}. Among them, authors in \cite{dzg1} propose to select two users with large differences on channel coefficients into one group to improve the overall system throughput of NOMA systems. This strategy is expanded to the scenario with more users in each group by Ali \emph{et al}. in \cite{Ali1}. Moreover, matching theory and matching game are also applied into user grouping\cite{Marcano2018, zhang1, Xu2018, Di1}. To increase the system capacity in half-duplex cognitive orthogonal frequency division multiplexing(OFDM)-NOMA systems Xu \emph{et al}. in \cite{Xu2018} propose a user grouping strategy depending on matching theory. To balance the weighted sum rate(WSR) and fairness among users, in \cite{Di1}, the authors assign users into different groups depending on matching theory.

Some researchers have focused on user grouping in multi-cell NOMA systems \cite{Networks2018k, Mokhtari2019k, Zhang2017k}. The authors in \cite{Zhang2017k} and \cite{Zhang2019a} expanded the strategy of selecting two users whose channel conditions are more distinctive into the same group in \cite{dzg1} to multi-cell NOMA systems and multi-cell downlink multiple-input multiple-output NOMA networks, respectively. In addition, joint optimization of user grouping and power allocation has been investigated. L. You \emph{et al} in \cite{Ding2019Two-side} derive a unified optimization framework for the joint optimization problem of user grouping and power allocation in multi-cell NOMA systems. {To maximize the download elastic traffic rate in multi-cell NOMA systems, a joint optimization strategy of subchannel assignment and power allocation was} proposed in \cite{Mokhtari2019k}. In these studies, the impact of inter-cell interference on SIC has not been well addressed. The SIC decoding order is still the same as that in single-cell systems, i.e., the increasing order of channel gains.

{It is still challenging for existing user grouping strategies to achieve the optimal performance, like power consumption, for multi-cell NOMA systems. First, in practice, QoS requirements, such as target data rate, of different users have also not been well concerned in the existing user grouping strategies in multi-cell NOMA systems. Diverse QoS requirements may impact the way to allocate power and group users. The reason is that, to satisfy the various QoS requirements of users, adequate power is needed, where more power will be allocated to the high target data rates users. It means that more inter-user interference will be produced by the users with higher target data rates.
Therefore, to alleviate inter-user interference in each group, it is important to avoid high target data rates users being assigned into the same group.
{If we group users considering diverse QoS requirements, total transmit power can be further reduced due to the alleviating of inter-user interference. It is also the meaning of ``QoS-aware'' in the title of this paper.}
In existing studies, this issue has not been addressed. Second, inter-cell interference is necessary to be considered in practical multi-cell NOMA systems. The efficiency of SIC will be seriously affected by inter-cell interference and inter-cell interference is different among different groups in each base station (BS). The widely used SIC decoding order of increasing channel gains in existing studies is no longer optimal, in which inter-cell interference is not well considered\cite{Yang2018,vaezi2019multiple}. Last, to reduce the complexity of user grouping, the number of users in each group is assumed to be equal in most existing user grouping strategies. If the number of users in each group is unfixed and arbitrary, a better user grouping strategy with less inter-user interference may be found.}

To address aforementioned challenges, in this paper, we study QoS-aware user grouping strategy that minimizes power consumption in multi-cell NOMA systems with inter-cell interference. Our preliminary work was presented in \cite{guo2019interference}. In \cite{guo2019interference}, we solved the user grouping problem in single-cell NOMA systems with QoS constraints using graph theory. In this paper, we extend our work in \cite{guo2019interference} to multi-cell NOMA systems with inter-cell interference.
{The novelty of this paper mainly lies in two aspects: 1) We focus on a novel QoS aware user grouping problem in a downlink multi-cell NOMA system, while most existing studies perform user grouping based on channel conditions. Furthermore, the optimal SIC decoding order is derived with consideration of diverse QoS requirements in this system, which is different from that based on channel conditions. 2) We propose a novel method to solve the problem from the perspective of graph theory. Particularly, we convert the problem of finding a grouping strategy into the problem of looking for some specific negative loops in the graph composed of users.} The main contributions are described as follows.
\begin{itemize}
\item Under different QoS requirements, we first analyze the optimal SIC decoding order and the optimal power allocation strategy for any given user grouping strategy considering downlink multi-cell NOMA scenario with inter-cell interference. Then we formulate the user grouping problem for power minimization in multi-cell NOMA systems.

\item The game model of the multi-cell user grouping problem is obtained, where each BS is a player with goal to find an action to reduce its action effect function. We prove that the obtained game model can be classified as an exact potential game. In addition, {the existence of the Nash Equilibrium(NE) and the finite improvement property (FIP) of the proposed game {are} also proved.}% A heuristic strategy is proposed to describe the dynamic procedure of this game.

\item To find the grouping strategy of each player in the obtained game, a weighted directed graph is constructed, and the problem of grouping users in each BS is converted into the problem of looking for some specific negative loops in this graph. Then we extend Bellman-Ford algorithm to find these negative loops to minimize total transmit power. To approach the optimal user grouping strategy within polynomial time, a fast greedy strategy is also proposed.

\end{itemize}

Extensive simulations are carried out to evaluate the performance of the proposed strategies. We first show that the proposed two strategies can converge to stable solutions after finite iterations. Then the traditional successive interference cancellation(SIC) decoding order of increasing channel gains is proved to be not optimal. By comparing with two reference user grouping strategies, we show that the proposed two user grouping strategies can significantly reduce inter-cell interference as well as total transmit power. Furthermore, the simulation results indicate that the proposed two strategies avoid users with high target data rate or poor channel condition being grouped into the same group as much as possible, resulting in reduction of inter-cell interference and total transmit power.

The rest of this paper is organized as follows. Section II outlines the model of the downlink multi-cell NOMA systems. We analyze and formulate the problem of user grouping for transmitting power minimization with QoS constraints in Section III. In Section IV, we prove that this problem can be formulated as an exact potential game. In Section V, we convert the action selection problem of each player, i.e., each BS, into the problem of looking for some specific negative loops in the graph composed of users and propose two strategies to solve it. Systems performance is evaluated in Section VI. Finally, Section VII concludes the paper.

\emph{Notations:} Bold and calligraphic letters denote vectors and sets, respectively. $\cup$ and $\setminus$ represent set union and set difference operators, respectively.
$\lceil \cdot\rceil $ denotes the ceiling function. $\|\mathcal{A}\|$ denotes the number of elements in $\mathcal{A}$. $\parallel\!\!\cdot\! \!\parallel_1$ is the 1-norm.
Table I lists the main notations used in this paper.

\section{Systems Model}

%All the BSs cooperate via a
{We consider a downlink multi-cell NOMA system with $M$ BSs and $N$ users as shown in Fig. $\ref{fig:01}$.
We assume that all BSs share the same spectrum, and all the BSs can communicate and coordinate with each other, similar to that in \cite{Zeng2019,Mokhtari2019k}.}
There are $G$ channels in each BS. Let $\!\mathcal{M}\!\!=\!\!\{1, \cdots \!, M\} $, $\mathcal{N}$ $\!=\!\{1, \cdots \!, N\} $ and $\mathcal{G}\!=
\!\{1,\cdots\!, G\}$ denote the index sets of BSs, users and channels, respectively. The users in the range of BS m ($m \in \mathcal{M}$) simultaneously associating channel $g $ ($g \in\mathcal{G}$) compose group $mg$. Perfect channel state information (CSI) is assumed to be available at BSs and users\cite{WangYazheng2021L}. The indexes set of users assigned into BS $m$ and group $mg$ are denoted by $\mathcal{U}^{m}$ and $\mathcal{U}^{m,g}$, respectively, where the number of users in $\mathcal{U}^{m,g}$ is $\|\mathcal{U}^{m,g}\|$. We assume that each user associates to the nearest BS \cite{Zhang2017k}.

\begin{table}
\setlength{\abovecaptionskip}{-0.04cm}
\setlength\abovedisplayskip{1pt}
\setlength\belowdisplayskip{1pt}
\fontsize{8}{12}\selectfont
	\centering
	\caption{Main Notations}
	\begin{tabular} {|m{30pt}|	m{180pt}|}
    \hline
		Symbol		& Description							 	\\
    \hline
		$ \mathcal{N} $ 		& Index set of all users 										\\	
    \hline
		$ \mathcal{G} $ 		& Index set of subchannels(groups) 									\\	
    \hline
		$ \mathcal{M} $ 		& Index set of BSs 									\\		
    \hline
		$ \mathcal{U}^{m,g} $ 		& Index set of users in group $ mg $ 							\\
    \hline
$ p_n $ 			& Transmit power allocated to user $ n $ 	\\
    \hline
		$ p^{m,g}_k $ 			& Transmit power allocated to the $ k $-th user in group $ mg $ 	\\
    \hline
		$ P_t $ 				& Total transmit power					\\
    \hline
		$ p^{m,g} $ 				& Total transmit power in group $ mg $		\\
    \hline
		$ \pi_n $ 				& Group that user $n$ assigned to 						\\
    \hline
		$ \bm{\pi} $ 			& Vector of grouping strategy of all users 			\\
    \hline
		$ \bm{\pi}_{-n} $ 		& Vector of grouping strategy of users except user $ n $ 	\\
    \hline
		$ (n\!\!\rightarrow\!\! g) $ 	& Operation that moving user $ n $ into group $ g $	\\
    \hline
        	$ (n\!\!\twoheadrightarrow\!\!\tilde{n}) $ & Operation that moving user $ n $ into the group including user $ \tilde{n} $, and moving user $ \tilde{n} $ out of its original group\\
    \hline
	\end{tabular}
	\label{table1}
\end{table}

\begin{figure*}[htbp]
	%\vspace{-0.3cm}
	\setlength{\abovecaptionskip}{-0.01cm}
	\setlength{\belowcaptionskip}{-0.4cm}
	\centering
	\setlength{\belowcaptionskip}{-16pt}
\setlength{\belowcaptionskip}{-1cm}
	\includegraphics[width=6in]{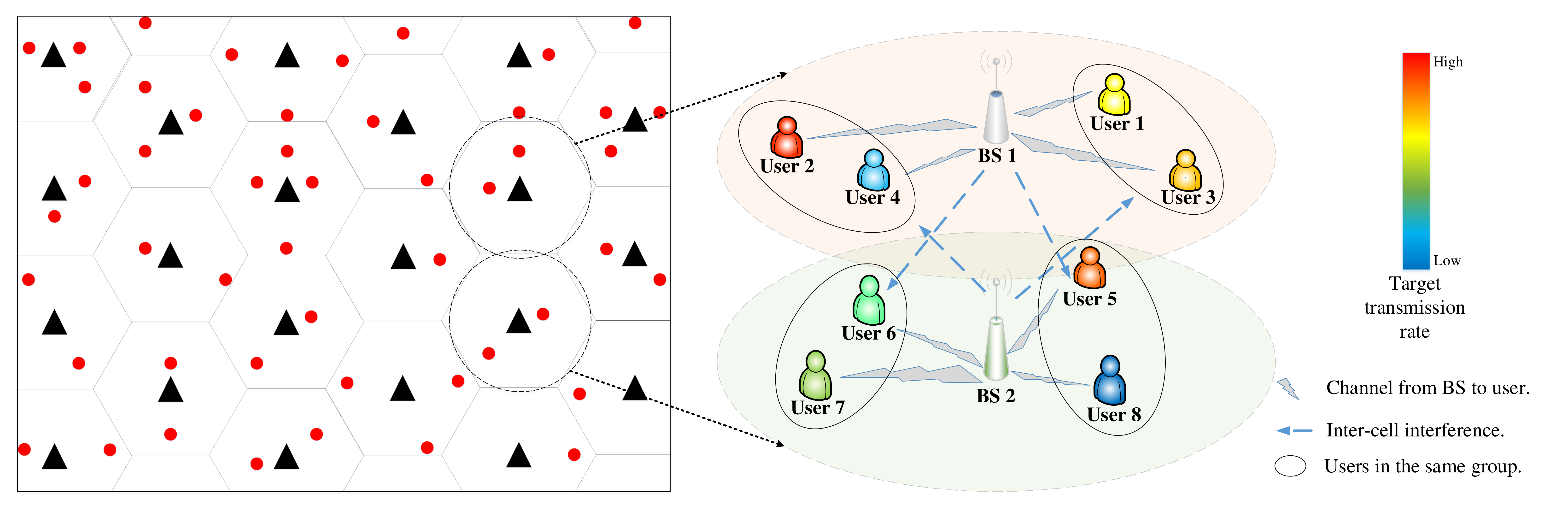}
	\caption{Illustration of Downlink Multi-cell  NOMA}
	\label{fig:01}
	%\floatsep
\end{figure*}

\begin{figure*}[htbp]%
	%\vspace{-0.3cm}
	\setlength{\abovecaptionskip}{-0.01cm}
	\setlength{\belowcaptionskip}{-1cm}
	\centering
	\setlength{\belowcaptionskip}{-16pt}
\setlength{\belowcaptionskip}{-1cm}
	\includegraphics[width=6in]{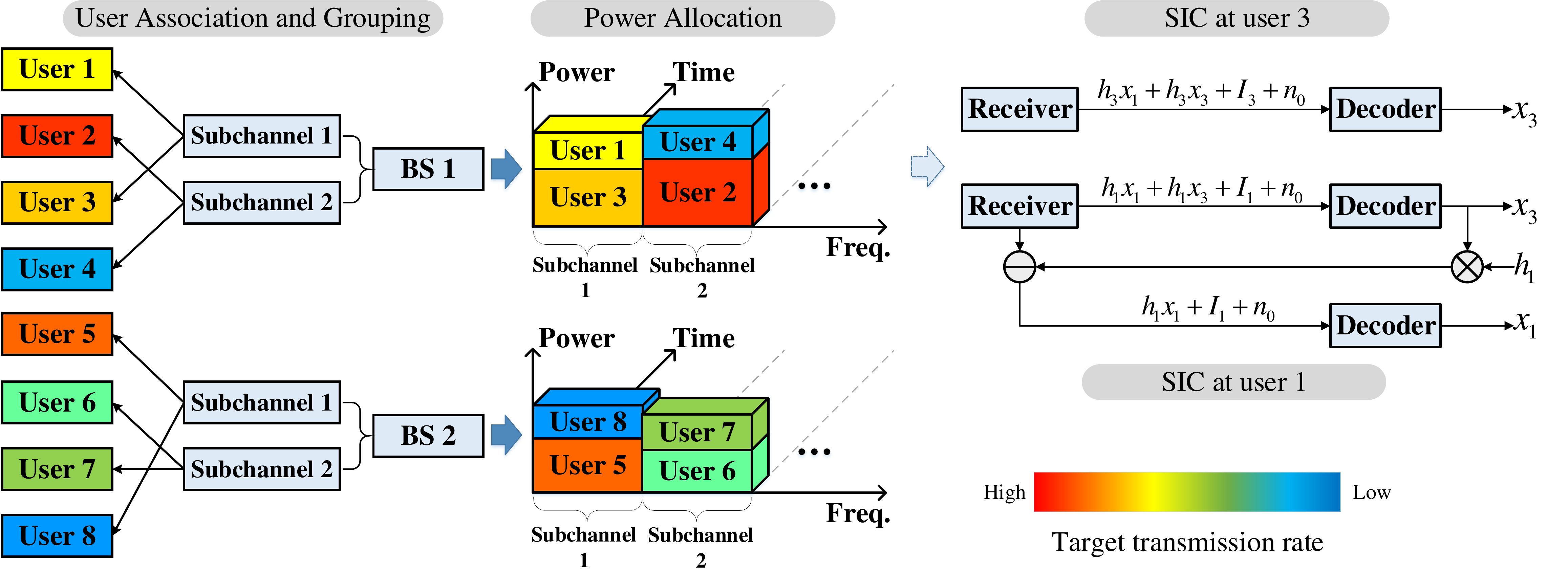}
	\caption{Process of Downlink Multi-cell  NOMA with SIC}
	\label{fig:001}
	%\floatsep
\end{figure*}

According to the NOMA principle, BS $m$ exploits SC and broadcasts signal $\sum_{n\in\mathcal{U}^{m,g}}\sqrt{p_{n}}s_n$ to users in group $mg$, {where $s_n$ denotes the transmitted symbol of user $n$, and $\mathbb{E}[s_n] = 0$ and $\mathbb{E}[| s_n|^2] = 1$ for $\forall n \in \mathcal{N}$.}
The signal that user $n$ receives is given by $y_n=H^{m,g}_{n}\sum_{i\in\mathcal{U}^{m,g}}\sqrt{p_{i}}s_i+\sum_{m'\in\mathcal{M}\setminus {m}}H^{m',g}_{n}\sum_{i\in\mathcal{U}^{m',g}}\sqrt{p_{i}}s_i+\mathfrak{n}_{n},~~n\in \mathcal{U}^{m,g}$, where $\mathfrak{n}_{n}\thicksim\mathcal{N}(0, \sigma^2)$ is Additive White Gaussian Noise (AWGN). $H^{m,g}_{n}$ denotes the channel coefficient between BS $m$ ($m \in \mathcal{M}$) and user $n $ ($n\in \mathcal{N}$) on subchannel $g $ ($g\in \mathcal{G}$).

The process of downlink multi-cell NOMA with SIC is presented in Fig. $\ref{fig:001}$. As shown in the upper right part of Fig. 1 and lower right part of Fig. 2, {the users with different
target transmission rates are} distinguished by their colors. In Fig. 1, the distances between users and BSs indicate their channel conditions. To make it more clear, each user in Fig. 2 is indexed with the same index with the user in the same color in Fig. 1. First, each user associates to the nearest BS \cite{Zhang2017k}.Then user association and grouping are executed based on their channel condition and target transmission rate. {BS allocates different transmit power to different users} and users distinguish received signals by SIC\cite{ZhangMing2021MC}. A simple process of SIC in downlink NOMA taking the case of user $1$ and user $3$ is shown at the right of Fig. $\ref{fig:001}$, where $x_1$ and $x_3$ are signals transmitted to user $1$ and user $3$ from BS $1$, respectively. $h_1$ and $h_3$ are channel coefficients between BS $1$ and user 1 and user 3, respectively.

{With SIC, each user needs to decode the signals of users with higher priority detecting order. Therefore, if $n,i\in\mathcal{U}^{m,g}$ and user $n$ has higher priority detecting order than user $i$, user $i$ needs to decode the signal of user $n$ before decoding its own signal. When user $i$ decodes the signal of user $n$, i.e., $\sqrt{p_{n}}s_n$, all the signals of users with lower detecting order than user $n$, i.e., $H_{i}^{m,g}\!\!\!\!\!\!\!\!\!\sum\limits_{_{S^{_{m,g}}_{j}>S^{m,g}_{n}}^{~~j\in\mathcal{U}^{m,g}} }\!\!\!\!\!\!\!\!\!\sqrt{p_{j}}s_j$, will be regarded as interference.
For convenience, according to the Shannon's theorem, the achievable data rate for user $i$ to decode the signal of user $n$ is denoted by $R_{i\dashrightarrow n}$ (particularly, $R_{n\dashrightarrow n}=R_{n}$), which is expressed as\cite{Yang2018}
%According to the Shannon's theorem, the data rate achievable of user $n\in \mathcal{U}^{m,g}$ ($n \in \mathcal{N}$,$m \in \mathcal{M}$, $g \in \mathcal{G}$) is given by:
\begin{equation}\small
\setlength\abovedisplayskip{1pt}
\setlength\belowdisplayskip{1pt}
\label{re_a601}
\begin{aligned}
R_{i\dashrightarrow n}\!\!= \!\!log_2
\Bigg(\!\!1\!\!+\!\!\tfrac{|H_{i}^{m,g}|^2p_{n}}{|H_{i}^{m,g}|^2\!\!\!\!\!\!\!\!\sum\limits_{_{S^{_{m,g}}_{j}>S^{m,g}_{n}}^{~~\!j\in\mathcal{U}^{m,g}} }\!\!\!\!\!\!\!\!p_{j}+I_i+\sigma^2}\!\!
 \Bigg),~~n,i\in\mathcal{U}^{m,g},%~~~~~n\in \mathcal{U}^g.
\end{aligned}
\end{equation}where $S^{m,g}_{n}$ is associated with the proposed SIC decoding order in Theorem 1 of the next section, and the user with lower $S^{m,g}_{n}$ has higher priority detecting order, $I_n$ denotes the total interference from the other BSs, i.e.,
{
	\begin{equation}\label{2000+}\small
\setlength\abovedisplayskip{1pt}
\setlength\belowdisplayskip{1pt}
	I_n=\!\!\!\sum\limits_{m'\in \mathcal{M}\setminus {m}}\!\!\!|H^{m',g}_{n}|^2\!\!\!\sum\limits_{j\in \mathcal{U}^{m',g}}\!\!\!\!p_{j},~~~n\!\in\! \mathcal{U}^{m,g}.
	\end{equation}}\quad By SIC, a user should correctly detect information of the users with higher priority detecting order in the same group\cite{Xia2018}. Therefore, to make sure that the information of one user can be detected by the users with lower priority detecting order, i.e., $R_{n} \!<R_{i\dashrightarrow n}, \forall i,n \in\mathcal{U}^{m,g}, S^{_{m,g}}_{i}> S^{m,g}_{n}$, we will have
{\begin{equation}\small
\setlength\abovedisplayskip{1pt}
\setlength\belowdisplayskip{1pt}
\label{2611}
\begin{aligned}
&R_{n} \!\!=\!\!\!\!\! \!\min\limits_{_{S^{_{m,g}}_{i}\geq S^{m,g}_{n}}^{~~i\in\mathcal{U}^{m,g}}}\!\!\!\!\!R_{i\dashrightarrow n}\\
\!=&\!\!\!\! \!\min\limits_{_{S^{_{m,g}}_{i}\geq S^{m,g}_{n}}^{~~i\in\mathcal{U}^{m,g}}}\!\! log_2
\Bigg(\!\!1\!+\!\tfrac{|H_{i}^{m,g}|^2p_{n}}{|H_{i}^{m,g}|^2\!\!\!\!\sum\limits_{_{S^{_{m,g}}_{j}>S^{m,g}_{n}}^{~~j\in\mathcal{U}^{m,g}} }\!\!\!\!\!p_{j}+I_i+\sigma^2}
 \Bigg),~~~n\!\in\! \mathcal{U}^{m,g}.\end{aligned}
\end{equation}}}To satisfy the QoS requirement of each user, the achievable data rate of a user should be no less than its target data rate:
{{
\begin{equation}\label{2612}\small
\setlength\abovedisplayskip{0pt}
\setlength\belowdisplayskip{0pt}
 R_n\geq r_{n},~~~~n\in \mathcal{N}.
\end{equation}}}
\quad Combine ($\ref{2611}$) with ($\ref{2612}$), the following constraint about power allocation is obtained.
{
\begin{equation*}\label{2613}\small
\setlength\abovedisplayskip{1pt}
\setlength\belowdisplayskip{1pt}
\min\limits_{_{S^{_{m,g}}_{i}\geq S^{m,g}_{n}}^{~~i\in\mathcal{U}^{m,g}}} log_2
\Bigg(\!1+\tfrac{|H_{i}^{m,g}|^2p_{n}}{|H_{i}^{m,g}|^2\!\!\!\!\!\!\!\sum\limits_{_{S^{_{m,g}}_{j}> S^{m,g}_{n}}^{~~j\in\mathcal{U}^{m,g}} }\!\!\!\!\!\!\!p_{j}+I_i+\sigma^2}
 \Bigg)\geq r_n,~~~~n\in \mathcal{N}.%\quad n\in \mathcal{U}^g, g\in \mathcal{G},
\end{equation*}}Then the minimal transmit power allocated to user $ n$ is expressed as follows.
{
\begin{equation}\label{2614}\small
\setlength\abovedisplayskip{0pt}
\setlength\belowdisplayskip{1pt}
p_{n}\geq(2^{r_{n}}-1)(\!\!\!\!\sum\limits_{_{S^{_{m,g}}_{i}< S^{m,g}_{n}}^{~~j\in\mathcal{U}^{m,g} }}\!\!\!\!\!p_{j}+\!\!\!\max\limits_{^{~~i \in \mathcal{U}^{m,g}}_{S^{_{m,g}}_{i}\geq S^{m,g}_{n}}}\!\!\!\tfrac{\sigma^2+I_i}{|H_{i}^{m,g}|^2}),~~~n\!\in\! \mathcal{U}^{m,g}.
\end{equation}}The total transmit power can be obtained as follows.
{\begin{equation}\label{2615}\small
\setlength\abovedisplayskip{0pt}
\setlength\belowdisplayskip{0pt}
P_t=\sum\limits_{m\in \mathcal{M}}\sum\limits_{g\in \mathcal{G}}\sum\limits_{n\in \mathcal{U}^{m,g}}p_{n}.
\end{equation}}

%\vspace{-0.5cm}
\section{Problem Formulation}\label{section3}% and the PCE Function in Downlink NOMA

{Based on the analysis above, the joint user grouping and power allocation problem for minimizing total transmit power in multi-cell NOMA systems is formulated as follows.{%\vspace{-0.2cm}
\begin{subequations}\label{3613}\small
\setlength\abovedisplayskip{1pt}
\setlength\belowdisplayskip{1pt}
	\begin{align}
	\min_{{\mathcal{U}}^{m,g},p_{n}} ~ &P_t=\sum\limits_{m\in \mathcal{M}}\sum\limits_{g\in \mathcal{G}}\sum\limits_{n\in \mathcal{U}^{m,g}}p_{n}\\
     \text{s.t.}~~~&p_{n}\!\!\geq\!(2^{r_{n}}\!\!-\!\!1)(\!\!\!\!\!\!\!\!\!\sum\limits_{_{S^{_{m,g}}_{i}< S^{m,g}_{n}}^{~~j\in\mathcal{U}^{m,g} }}\!\!\!\!\!\!\!\!p_{j}\!+\!\!\!\!\max\limits_{^{~~i \in \mathcal{U}^{m,g}}_{S^{_{m,g}}_{i}\geq S^{m,g}_{n}}}\!\!\!\!\!\!\tfrac{\sigma^2+I_i}{|H_{i}^{m,g}|^2}),~~n\!\in\! \mathcal{U}^{m,g},\\
     &\bigcup\limits_{m\in \mathcal{M}}\bigcup\limits_{g\in \mathcal{G}}{\mathcal{U}^{m,g}}=\mathcal{N},\\
    &{\mathcal{U}^{m,g}}\cap{\mathcal{U}^{m,g'}}=\varnothing,~~~g,g'\in \mathcal{G},g\neq g',\\
    &{\mathcal{U}^{m,g}}\cap{\mathcal{U}^{m',g}}=\varnothing,~~~m,m'\in \mathcal{M},m\neq m'.
	\end{align}
\end{subequations}}where ($\ref{3613}$c), ($\ref{3613}$d) and ($\ref{3613}$e) mean that every user should be assigned into one group.}

Since the radio resources among different subchannels are orthogonal, there is no inter-user interference among different subchannels. Furthermore, the minimal power consumption of each user in $\mathcal{N}$ can be obtained by ($\ref{2614}$). But ($\ref{2614}$) is very complicated due to the maximization function. To simplify ($\ref{2614}$), the optimal SIC decoding order is required. However, due to serious inter-cell interference, the existing SIC decoding order that sorting users by the increasing order of channel gains in single-cell system is not optimal in multi-cell system\cite{Yang2018}. Therefore, we propose a novel SIC decoding order for multi-cell system in Theorem $\ref{thm1-}$. For simplification, we define a channel coefficient to interference plus noise ratio(CCINR) as
{
	\begin{equation}\label{3611}\small
\setlength\abovedisplayskip{1pt}
\setlength\belowdisplayskip{1pt}
	S^{m,g}_n=\tfrac{|H^{m,g}_{n}|^2}{I_n+\sigma^2}, ~~~~n\in \mathcal{U}^{m,g}.
	\end{equation}}
\begin{thm1}
\label{thm1-}
To minimize total transmit power while satisfying the QoS requirement of each users, the optimal SIC decoding order is the ascending order of $S^{m,g}_n$ and we should allocate power according to ($\ref{3612}$) to each user in group $mg$ in descending order of $S^{m,g}_n$.
{
\setlength\abovedisplayskip{0pt}
\setlength\belowdisplayskip{0pt}
		\begin{equation}\label{3612}\small
\begin{aligned}
p_n&=(2^{r_{n}}-1)(\tfrac{I_n+\sigma^2}{|H^{m,g}_{n}|^2}+\!\!\!\!\!\!\!\sum\limits_{_{S^{_{m,g}}_{i}>S^{m,g}_{n}}^{~~\!i\in\mathcal{U}^{m,g}}}\!\!\!\!\!\!p_{i}),~~~~n\in \mathcal{N}.
\end{aligned}
		\end{equation}}
\end{thm1}
\vspace{-0.2cm}
\begin{proof}
\vspace{-0.2cm}
See Appendix A.\qedhere
\end{proof}

Let $p^{m,g}$ denote total power consumption of users in group $mg$, i.e.,
{\setlength\abovedisplayskip{0pt}
\setlength\belowdisplayskip{0pt}
		\begin{equation}\label{3614}\small
\begin{aligned}
&p^{m,g}\!\!=\!\!\!\!\!\sum\limits_{n\in \mathcal{U}^{m,g}}\!\!\!\!\!p_{n}%=&\sum\limits_{n\in \mathcal{U}^{m,g}}(2^{r_{n}}-1)(\tfrac{I_n+\sigma^2}{|H^{m,g}_{n}|^2})\prod\limits_{_{S^{_{m,g}}_{i}<S^{m,g}_{n}}^{~~i\in\mathcal{U}^{m,g}}}2^{r_{i}}\\
\!=\!\!\!\!\!\!\sum\limits_{n\in \mathcal{U}^{m,g}}\!\!\!\!(2^{r_{n}}\!\!-\!\!1)(\tfrac{\sigma^2}{|H^{m,g}_{n}|^2})\!\!\!\!\!\!\!\!\prod\limits_{_{S^{_{m,g}}_{i}<S^{m,g}_{n}}^{~~i\in\mathcal{U}^{m,g}}}\!\!\!\!\!\!\!\!\!2^{r_{i}}\!\!\\
+&\!\!\!\!\!\!\!\!\!\sum\limits_{m'\in \mathcal{M}\setminus \{m\}}\!\!\!\!\!\!\!\!\!p^{m',g}\!\!\!\sum\limits_{n\in
\mathcal{U}^{m,g}}\!\!\!\!(|H^{_{m',g}}_{n}|^2)(\tfrac{2^{r_{n}}-1}{|H^{m,g}_{n}|^2})\!\!\!\!\!\!\!\!\prod\limits_{_{S^{_{m,g}}_{i}<S^{m,g}_{n}}^{~~i\in\mathcal{U}^{m,g}}}\!\!\!\!\!\!\!\!2^{r_{i}}.
\end{aligned}
\end{equation}}The detailed derivation process of ($\ref{3614}$) is proposed in Appendix B.

%@@@@@@@@@@@@@@@@@@@@@@
%
%An interesting strategy.

Let $\mathbb{A}^g$ denote a $M\times M$ matrix, where
{
		\begin{equation}\label{3615}\small
\setlength\abovedisplayskip{1pt}
\setlength\belowdisplayskip{1pt}
\begin{aligned}
\mathbb{A}^g_{mm'}\!=\!\!\left\{\begin{array}{ll}
\!\!\sum\limits_{n\in
\mathcal{U}^{m,g}}\!\!(|H^{_{m',g}}_{n}|^2)(\tfrac{2^{r_{n}}-1}{|H^{m,g}_{n}|^2})\!\!\!\!\!\!\!\prod\limits_{_{S^{_{m,g}}_{i}<S^{m,g}_{n}}^{~~i\in\mathcal{U}^{m,g}}}\!\!\!\!\!\!2^{r_{i}},&m\neq m'\\
\!\!-1,&m= m'\\
 \end{array}\right..
\end{aligned}
\end{equation}}Let $\mathbf{b}$ denote a $M\times 1$ vector, where
{
		\begin{equation}\label{3616}\small
\setlength\abovedisplayskip{1pt}
\setlength\belowdisplayskip{1pt}
\begin{aligned}
\mathbf{b}^g_{m}=-\sum\limits_{n\in \mathcal{U}^{m,g}}(2^{r_{n}}-1)(\tfrac{\sigma^2}{|H^{m,g}_{n}|^2})\!\!\!\!\!\!\prod\limits_{_{S^{_{m,g}}_{i}<S^{m,g}_{n}}^{~~~i\in\mathcal{U}^{m,g}}}\!\!\!\!\!\!2^{r_{i}}.\\
\end{aligned}
\end{equation}}Let $\mathbf{P}^g$ denote a $M\times 1$ vector, where $\mathbf{P}^g_m=p^{m,g}$. Then according to ($\ref{3614}$), we have:
{\setlength\abovedisplayskip{1pt}
\setlength\belowdisplayskip{1pt}
		\begin{equation}\label{3617}\small
\begin{aligned}
\mathbb{A}^g\mathbf{P}^g=\mathbf{b}^g \Longrightarrow \mathbf{P}^g=\overline{\mathbb{A}^g}\mathbf{b}^g.\\
%&\Downarrow\\
%\mathbf{P}^g&=\overline{\mathbb{A}^g}\mathbf{b}^g
\end{aligned}
\end{equation}}where $\overline{\mathbb{A}^g}$ is the inverse matrix of $\mathbb{A}^g$. The reversibility of matrix $\mathbb{A}^g$ is proved in Appendix C. Then {the} total power
consumption of all users on channel $g$ is:
{
		\begin{equation}\label{3618}\small
\setlength\abovedisplayskip{1pt}
\setlength\belowdisplayskip{1pt}
\begin{aligned}
\sum\limits_{m\in \mathcal{M}}p^{m,g}=\parallel\mathbf{P}^g\parallel_1&=\parallel\overline{\mathbb{A}^g}\mathbf{b}^g\parallel_1.
\end{aligned}
\end{equation}}where $\parallel\cdot \parallel_1$ is the 1-norm.

{Since the minimal transmit power can be calculated through Theorem \ref{thm1-} for any user grouping strategies, what we need to do next is to find the optimal user grouping strategy under Theorem \ref{thm1-} to minimize total transmit power. }Therefore, problem ($\ref{3613}$) can be rewritten as follows.
{
\setlength\abovedisplayskip{1pt}
\setlength\belowdisplayskip{1pt}
\begin{subequations}\label{3619}\small
	\begin{align}
	\min_{{\mathcal{U}}^{m,g}} ~~ &P_t=\sum\limits_{m\in \mathcal{M}}\sum\limits_{g\in \mathcal{G}}p^{m,g}\\
     \text{s.t.}~~~~&\sum\limits_{m\in \mathcal{M}}p^{m,g}=\parallel\overline{\mathbb{A}^g}\mathbf{b}^g\parallel_1,\\
     &\mathbb{A}^g_{mm'}\!\!=\!\left\{\begin{array}{ll}
\!\!\!\!\sum\limits_{n\in
\mathcal{U}^{m,g}}\!\!\!\!(|H^{_{m',g}}_{n}|^2)(\tfrac{2^{r_{n}}-1}{|H^{m,g}_{n}|^2})\!\!\!\!\!\!\!\!\!\!\prod\limits_{_{S^{_{m,g}}_{i}<S^{m,g}_{n}}^{~~~\!i\in\mathcal{U}^{m,g}}}\!\!\!\!\!\!\!\!\!2^{r_{i}},&m \neq m'\\
-1,&m= m'
 \end{array}\right.,\\
 &\mathbf{b}^g_{m}=-\!\!\!\!\sum\limits_{n\in
 \mathcal{U}^{m,g}}\!\!\!\!(2^{r_{n}}-1)(\tfrac{\sigma^2}{|H^{m,g}_{n}|^2})\!\!\!\!\!\!\!\!\!\prod\limits_{_{S^{_{m,g}}_{i}<S^{m,g}_{n}}^{~~~\!i\in\mathcal{U}^{m,g}}}\!\!\!\!\!\!2^{r_{i}}\\
     &\bigcup\limits_{m\in \mathcal{M}}\bigcup\limits_{g\in \mathcal{G}}{\mathcal{U}^{m,g}}=\mathcal{N},\\
    &{\mathcal{U}^{m,g}}\cap{\mathcal{U}^{m,g'}}=\varnothing,~~~g,g'\in \mathcal{G},g\neq g',\\
    &{\mathcal{U}^{m,g}}\cap{\mathcal{U}^{m',g}}=\varnothing,~~~m,m'\in \mathcal{M},m\neq m'.
	\end{align}
\end{subequations}}\quad Since the number of users that we assigned into each group is unfixed, the existing user grouping strategies based on matching theory or the matching game are no longer applicable. Besides, inter-cell interference is not considered in existing strategies, which has severe impact on NOMA transmission.
This problem is a mixed-integer non-linear programming (MINLP) problem whose feasible region is non-convex. The solution to such problem generally yields unacceptable computation burden by exhaustive search or similar algorithms. To solve this problem practically, we model it as an exact potential game where each BS is a player. We also propose an expanded Bellman-Ford algorithm and a greedy based suboptimal algorithm for each player to group users with lower computational complexity. The details of the proposed modeling and algorithms are presented in Section IV and Section V, respectively.

\section{Potential Game}

We define a subchannel assigning vector $\bm{\pi}=[\pi_{n}]_{1\leq n \leq N,}$, where $ \pi_n $ is the subchannel index of user $ n $. Let $\bm{\pi}_{-n}$ denote the vector of subchannel indexes of users in $\mathcal{N}\!\setminus\!\{n\}$, i.e., $\bm{\pi}_{-n}\!\!=\!\!(\pi_1\!, \cdots\!\!, \pi_{n-1}, \pi_{n+1},\cdots, \pi_{N})$.
Let $(n\!\!\!\rightarrow\!\! g)$ denote the operation that user $n$ is moved into the group served by subchannel $g$ in the same BS. Similarly, $(n\!\!\rightarrow\!\! \pi_{\tilde{n}})$ denotes the operation that user $n$ is moved into the group containing user $\tilde{n}$.

{We assume that all users associated to BS $m$ have been numbered from $1$ to $\|\mathcal{U}^{m}\|$. We assume that user $m_k$ is the $k$-th user in BS $m$. Then the group containing the $k$-th user in BS $m$ can be expressed as ${\pi}_{m_k}$. {To describe the action of all users in BS $m$ when the user grouping strategy of BS $m$ changes}, a $\|\mathcal{U}^{m}\|\times G$ action matrix $\mathbb{J}^{m}$ is introduced, where $\mathbb{J}^{m}_{kg}$ is the element of this matrix at the $k$ row and the $g$ column. $\mathbb{J}^{m}_{kg}=1$ means that user $m_k$ {is} not in group $g$ and user $m_k$ should be moved into group $g$, i.e., action $(m_k\rightarrow g$), and $\mathbb{J}^{m}_{kg}=0$ otherwise. With initial grouping strategy $\bm{\pi}$, this definition can be expressed by the following formula:

\begin{equation}\label{3620}\small
\setlength\abovedisplayskip{1pt}
\setlength\belowdisplayskip{1pt}
\begin{aligned}
\mathbb{J}^{m}_{kg}=\!\left\{\begin{array}{ll}
1&(m_{k}\!\!\rightarrow\!\! g),{\pi}_{m_{k}}\neq g\\
0&Otherwise\\
 \end{array}\right.
\end{aligned} 1\leq k \leq \|\mathcal{U}^{m}\|
\end{equation}
where ${\pi}_{m_{k}}\neq g$ means the $k$-th user associated to BS $m$ {is} not in group $g$ with grouping strategy $\bm{\pi}$ before $\mathbb{J}^{m}_{kk'}$ being executed. Then any change in the grouping strategy of users served by BS $m$ can be presented by $\mathbb{J}^{m}$. For instance, we assume that there are 5 users associated to BS $m$, with initial user grouping strategy $\bm{\pi}$, these users are assigned into 3 groups: {users $m_1$, $m_2$ in group 1}, users $m_3$, $m_4$ in group 2,users $m_5$ in group 3. If the grouping strategy $\bm{\pi}$ is changed to $\bm{\pi}'$ by moving user 1 into group 2, moving user 3 into group 3 and moving user 5 into group 1, then the action matrix $\mathbb{J}^{m}$ can be written as $\mathbb{J}^{m}=$$\begin{bmatrix} 0 & 1 & 0 \\ 0 & 0 & 0 \\0 & 0 & 1 \\ 0 & 0 & 0\\ 1 & 0 & 0\end{bmatrix}$.}

Let $\bm{\pi}\!\!\xrightarrow[]{\mathbb{J}^{m}}\!\!\bm{\pi}'$ indicate that $\bm{\pi}$ can be changed into $\bm{\pi}'$ by action $\mathbb{J}^{m}$. We define an action effect function
$\mathcal{D}_m(\bm{\pi}\!\!\xrightarrow[]{\mathbb{J}^{m}}\!\!\bm{\pi}')$ as the difference of total transmit power after $\bm{\pi}\!\!\xrightarrow[]{\mathbb{J}^{m}}\!\!\bm{\pi}'$, i.e.,
{{%\setlength\abovedisplayskip{1pt}
		\begin{equation}\label{3621}\small
\setlength\abovedisplayskip{1pt}
\setlength\belowdisplayskip{1pt}
\begin{aligned}
&\mathcal{D}_m(\bm{\pi}\xrightarrow[]{\mathbb{J}^{m}}\bm{\pi}')
=P_{total}\Big|(\bm{\pi}')-P_{total}\Big|({\bm{\pi}})\\
=&\sum\limits_{g\in \mathcal{G}}\sum\limits_{m\in \mathcal{M}}p^{m,g}\Big| (\bm{\pi}')-\sum\limits_{g\in \mathcal{G}}\sum\limits_{m\in \mathcal{M}}p^{m,g}\Big| (\bm{\pi})\\
=&\sum\limits_{g\in \mathcal{G}}\|\overline{\mathbb{A}^g}\mathbf{b}^g\|_1\Big| (\bm{\pi}')-\sum\limits_{g\in \mathcal{G}}\parallel\overline{\mathbb{A}^g}\mathbf{b}^g\parallel_1\Big | (\bm{\pi})\\
%=&(2^{r_{i}}-1)\!\!\!\!\!\!\!\!\sum\limits_{^{~~~~i\in\mathcal{U}^{m,g}}_{{S^{_{m,g}}_{i}>S^{m,g}_{n}}}}\!\!\!\!\!\!\!\!(|H^{_{m',g}}_{i}|^2)(\tfrac{2^{r_{i}}-1}{|H^{m,g}_{i}|^2})\!\!\!\!\!\!\!\!\!\!\!\!\!\prod\limits_{_{|S^{_{m,g}}_{j}|<|S^{m,g}_{i}|}^{~~~~~\!i\in\mathcal{U}^{m,g}\setminus\{n\}}}\!\!\!\!\!\!\!\!\!\!\!\!2^{r_{j}}\\
%&+(|H^{_{m',g}}_{n}|^2)(\tfrac{2^{r_{n}}-1}{|H^{m,g}_{n}|^2})\!\!\!\!\!\!\!\!\!\!\!\!\!\prod\limits_{_{S^{_{m,g}}_{i}<S^{m,g}_{n}}^{~~~~~\!i\in\mathcal{U}^{m,g}}}
\end{aligned}
\end{equation}}It should be noted that $\bm{\pi}=[\pi_{n}]_{1\leq n \leq N,}$contains the group information of all users, where $ \pi_n $ is the subchannel index of user $ n $. As mentioned in ($\ref{3618}$), $\sum\limits_{m\in \mathcal{M}}p^{m,g}=\parallel\overline{\mathbb{A}^g}\mathbf{b}^g\parallel_1$. $\|\overline{\mathbb{A}^g}\mathbf{b}^g\|_1\Big| (\bm{\pi}')=\sum\limits_{m\in \mathcal{M}}p^{m,g}\Big| (\bm{\pi}')$ and $\parallel\overline{\mathbb{A}^g}\mathbf{b}^g\parallel_1\Big | (\bm{\pi})=\sum\limits_{m\in \mathcal{M}}p^{m,g}\Big| (\bm{\pi})$ mean the total transmit power of users assigned with channel $g$ in condition of user grouping strategies $(\bm{\pi}')$ and $(\bm{\pi})$, respectively.} Set each BS as a player whose objective is to minimize its action effect function $\mathcal{D}_m(\bm{\pi}\xrightarrow[]{\mathbb{J}^{m}}\bm{\pi}')$ as defined in ($\ref{3621}$).
Accordingly, the game for this problem is defined as the tuple described as follows.
\begin{equation}\label{3622}%\nonumber
\small
\setlength\abovedisplayskip{1pt}
	\setlength\belowdisplayskip{1pt}
\mathscr{G}=\Big(\mathcal{M},\bm{\pi},\mathbb{J}^{m},\mathcal{D}_m(\bm{\pi}\xrightarrow[]{\mathbb{J}^{m}}\bm{\pi}')\Big),
\end{equation}where $\mathcal{M}$ is the set of players, $\bm{\pi}$ is the current grouping strategy. Game $\mathscr{G}$ is a pure strategic game, which means that the players' strategies are deterministic but not regulated by probabilistic rules \cite{Wang2010}. The NE is a strategy profile where no player has incentive to deviate unilaterally \cite{Liu2016}. The NE and the exact potential function in game $\mathscr{G}$ is defined as follows:
{\begin{def1}\label{def11}	
A user subchannel assigning vector $\bm{\pi}^*\!=\!(\pi_1^*, \ldots,\pi_N^*)$ is a NE if, for $\forall m \in \mathcal{M}$, we have:
\begin{equation}\label{3623}%\nonumber
\setlength\abovedisplayskip{1pt}
	\setlength\belowdisplayskip{1pt}
\small
\mathcal{D}_m(\bm{\pi}^*\xrightarrow[]{\mathbb{J}^{m}}\bm{\pi}')\geq0.
\end{equation}
\end{def1}
\vspace{-0.2cm}
This is, no BS has incentive to change its grouping strategy when $\bm{\pi}=\bm{\pi}^*$.}

\begin{def1}\label{def22}
%\vspace{-0.15cm}
$\Phi_m$ is an exact potential function of game $\mathscr{G}$, if for $\forall m \in\mathcal{M}$ and any given $\bm{\pi}$,  we have:
\begin{equation}\label{3624}\small%\nonumber
\setlength\abovedisplayskip{1pt}
\setlength\belowdisplayskip{1pt}
\begin{aligned}
\Phi_m(\bm{\pi}\!\!\xrightarrow[]{\mathbb{J}^{m}}\!\bm{\pi}')\!\!-\!\Phi_m(\bm{\pi}\!\!\xrightarrow[]{\tilde{\mathbb{J}}^{m}}\!\tilde{\bm{\pi}}')\!=\!P_{\!total}\Big|\!(\!\bm{\pi}')\!\!-\!\!P_{\!total}\Big|\!(\tilde{\bm{\pi}}').
\end{aligned}
\end{equation}
\end{def1}

Definition $\ref{def22}$ means that the difference of single player(each BS)'s utility due to unilaterally strategy deviation can exactly indicate the difference of total transmit power. In the following, we prove that game $\mathscr{G}$ is an exact potential game.

\begin{thm1}\label{thm2}
%\vspace{-0.15cm}
%\setlength{\abovecaptionskip}{-0.1cm}
\setlength{\belowcaptionskip}{-1cm}
Game $\mathscr{G}$ admits an exact potential function, which is expressed as follows:
\begin{equation}\label{3625}\small%\nonumber
\setlength\abovedisplayskip{1pt}
	\setlength\belowdisplayskip{1pt}
\begin{aligned}
 \Phi_m(\bm{\pi}\xrightarrow[]{\mathbb{J}^{m}}\bm{\pi}')=\mathcal{D}_m(\bm{\pi}\xrightarrow[]{\mathbb{J}^{m}}\bm{\pi}').\\
\end{aligned}
\end{equation}
\end{thm1}
\begin{proof}

\begin{equation}\small\nonumber\label{4008}
\begin{aligned}
\setlength\abovedisplayskip{1pt}
\setlength\belowdisplayskip{1pt}
&\mathcal{D}_m(\bm{\pi}\xrightarrow[]{\mathbb{J}^{m}}\bm{\pi}')-\mathcal{D}_m(\bm{\pi}\xrightarrow[]{\tilde{\mathbb{J}}^{m}}\tilde{\bm{\pi}}')\\
=&\Big(P_{total}\Big|(\bm{\pi}')-P_{total}\Big|{\bm{\pi}})\Big)-\Big(P_{total}\Big|(\tilde{\bm{\pi}}')-P_{total}\Big|({\bm{\pi}})\Big)\\
=&P_{total}\Big|(\bm{\pi}')-P_{total}\Big|(\tilde{\bm{\pi}}').
\end{aligned}
\end{equation}which satisfies ($\ref{3624}$) in Definition $\ref{def22}$. Therefore, the proof of Theorem $\ref{thm2}$ is concluded.
\qedhere
\end{proof}

Accordingly, game $\mathscr{G}$ is an exact potential game and admits an exact potential function
$\mathcal{D}_m(\bm{\pi}\xrightarrow[]{\mathbb{J}^{m}}\bm{\pi}')$. Then if we reduce the value of $\mathcal{D}_m(\bm{\pi}\xrightarrow[]{\mathbb{J}^{m}}\bm{\pi}')$ by action $\mathbb{J}^{m}$, total
transmit power will also reduce. Based on this, we prove the existence the FIP of this game as follows.

{\begin{cor1}\label{cor11}
	Game $\mathscr{G}$ is bound to stop at a NE, and has the FIP.
\end{cor1}
\begin{proof}	
	The FIP is a property for the game that is bound to stop after a limited number of iterations \cite{DM1}. It should be noted that every finite ordinal exact potential game has the FIP \cite{DM1} and there is at least one NE \cite{Liu2016} in a finite exact potential game\cite{Liu2016}. $\mathcal{G}$, $\mathcal{M}$ and $\mathcal{N}$ are all finite, and every time we change a BS's grouping strategy to reduce the value of its exact potential function, total transmit will also reduce. Thus a grouping strategy profile will not be selected repeatedly. So game $\mathscr{G}$ is bound to stop at a NE after a limited number of iterations. So the proof of Corollary 1 is concluded.\qedhere
\end{proof}}
\vspace{-0.3cm}

\subsection{Algorithm Description}

Based on the above analysis, the detailed game procedure is described in Algorithm $\ref{alg:1}$. It is a dynamic potential game strategy with complete information. A dynamic game is a game where players make decisions, i.e., choosing a grouping strategy, one by one but not in the meantime. Complete information means that all players know the strategies of the other players. {As mentioned in section II, all the BSs can communicate and coordinate with each other \cite{Zeng2019,Mokhtari2019k}.} Thus in Algorithm 1, the grouping strategy of each BS is chosen one by one according to the strategies of the other BSs.

{
%\vspace{-0.3cm}
\begin{algorithm}
\setlength\abovedisplayskip{1pt}
\setlength\belowdisplayskip{1pt}
	%\label{alg:1}
	\caption{Potential game based user grouping strategy}\label{alg:1}
	\KwIn{Set of users $\mathcal{N}$, set of BSs $\mathcal{M}$,
		target transmission rate $ {r}_n$, channel coefficient $ {H}^{m,g}_{n}$, ${n\!\!\in\!\mathcal{N}}$}
	\KwOut{Power ${p}_n, n\in\mathcal{N}$, Grouping Strategy $\bm{\pi}$}
	%\textbf{initialization:} Let the iteration index $k=1$.
    %\textbf{Initialize:}
	Set $\mathcal{U}^{m,g}=\varnothing, \forall g\in\mathcal{G}, m\in\mathcal{M}$;

	\For{$n=1 \textrm{ to }N$}{
		%Find $g_{min}=arg\min\limits_{^{g\in\mathcal{G}}_{n\in\mathcal{U}^g}}\big(|h_{n}|_{n\in\mathcal{U}^g}\big)$;
        Find $g_{max}=arg\max\limits_{^{g\in\mathcal{G}}}\big(|H_{n}^{m,g}|_{n\in\mathcal{U}^{m,g}}\big)$;

		%$\mathcal{U}^{g_{min}}\leftarrow\mathcal{U}^{g_{min}}\cup \{n\}$;
		 $\pi_n=g_{max}$;}
$\bm{\pi}=({\pi}_n)_{n\in \mathcal{N}}$;

\For{$n=1 \textrm{ to }N$}{
		Calculate $p_{n}$ according to ($\ref{3612}$).
	}
%$\mathcal{U}=\{\mathcal{U}^g\}_{g\in\mathcal{G}}$;
%$\mathcal{U}^g, \forall g\in\mathcal{G}$ and

	\Repeat{$\bm{\pi}$ doesn't change}{
		 %$Flag=0$;
		\For{$m=1 \textrm{ to }M$}{
         %\mathcal{D}_m(\bm{\pi}^*\xrightarrow[]{\mathbb{J}^{m}}\bm{\pi}')\geq0
         Find $\mathbb{J}^{m}\Big|_{\mathcal{D}_m(\bm{\pi}\xrightarrow[]{\mathbb{J}^{m}}\bm{\pi}')<0}$;

        Change user grouping strategy according to action $\mathbb{J}^{m}$;

        Update $p_{i}$ according to ($\ref{3612}$);
        }

		}
	\textbf{return} ${p}_n, n\in\mathcal{N}$, $\bm{\pi}$

\end{algorithm}
}
%\vspace{-0.3cm}

Once a user connects/disconnects a BS, what we need to do is to continue the game process based on the original strategy but not re-execute it. In the next section, the grouping strategy of each BS is analyzed by graph theory.

\section{Directed Graph for User Grouping in Downlink Multi-cell NOMA}

In this section, we will solve the problem that how to find an appropriate action $\mathbb{J}^{m}$ for each player, i.e., each BS. Based on the model of game $\mathscr{G}$, a weighted directed graph is builded, and the problem that how to find an appropriate strategy for each player, i.e., each BS{,} is solved based on graph theory. {According to the game process as described in Algorithm 1, we need to find action $\mathbb{J}^{m}\big|_{\mathcal{D}_m(\bm{\pi}\xrightarrow[]{\mathbb{J}^{m}}\bm{\pi}')<0}$. First we divide the actions into two types, \emph{shift league} and \emph{exchange league}:

\begin{def1}\label{defe1}
For any $i$ users indexed by $\{n_1,\cdots ,n_i\}$ in different groups of BS $m\in\mathcal{M}$,
if the subchannel assigning vector $\bm{\pi}$ is changed into $\tilde{\bm{\pi}}$ by $\mathbb{J}^{m}\big|\{(n_1\rightarrow\pi_{n_2}),\cdots ,(n_{i-1}\rightarrow \pi_{n_i})\}$, and total transmit power satisfies $\mathcal{D}_m(\bm{\pi}\xrightarrow[]{\mathbb{J}^{m}}\tilde{\bm{\pi}})<0 $, these $ i $ users compose an \emph{i-shift league}.
% \begin{equation*}
% \label{3005+}
% P_{t}(\tilde{\bm{\pi}}) < P_{t}(\bm{\pi})
% \end{equation*}
\end{def1}

%Where $(n\!\!\rightarrow\!\! g)$ stand the operation that user $n$ is moved into group $g$. Similarly, $(n\!\!\rightarrow\!\! \pi_{\tilde{n}})$ stand the operation that user $n$ is moved into the group including user $\tilde{n}$.

\begin{def1}\label{defe2}
For any $i$ users indexed by $\{n_1,\!\cdots \!,n_i\}$ in different groups of BS $m\in\mathcal{M}$. If the subchannel assigning vector $\bm{\pi}$ is changed into $\tilde{\bm{\pi}}$ by $\mathbb{J}^{m}\big|\{(n_1\!\!\rightarrow \!\!\pi_{n_2}),\cdots,(n_{i-1}\!\rightarrow \!\pi_{n_i}),(n_i\!\!\rightarrow \!\!\pi_{n_1})\}$, and total transmit power satisfies $
\mathcal{D}_m(\bm{\pi}\!\xrightarrow[]{\mathbb{J}^{m}}\!\tilde{\bm{\pi}})\!<\!0  $, these $ i $ users compose an \emph{i-exchange league}.
\end{def1}

Based on the definitions of \emph{shift league} and \emph{exchange league}, \emph{all-stable solution} can be defined as follow.

\begin{def1}\label{defe3}
A user grouping strategy based on subchannel assigning vector $\bm{\pi}$ is called \emph{all-stable solution}, if for $ \forall i \in \mathcal{N} $, there exists no \emph{i-shift league} or \emph{i-exchange league} in any BS.
\end{def1}

To explain the concept of ``all-stable solution'', we first introduce the concept of ``stable matching solution'' from \cite{Ronn1990}. A solution is called a stable matching solution if the overall gains of all agents cannot be improved by exchanging the assigned mates of any two agents. In this paper, a solution is a stable matching solution if there is no 2-\emph{exchange league} in it. Then we extend the concept of ``stable matching solution'' to ``all-stable solution'', which means that the overall gains of all agents cannot be improved by exchanging or shifting the assigned mates of any agents.

Although Definitions 3 and 4 are only for the case of one BS, \emph{all-stable solution} in Definitions 5 is defined under multi-cell cases. Then $\mathbb{J}^{m}\!\big|_{\!\mathcal{D}_m\!(\!\bm{\pi}\xrightarrow[]{\mathbb{J}^{m}}\bm{\pi}'\!)<0}$ in Algorithm $\ref{alg:1}$ can be found by searching for \emph{i-shift league} or \emph{i-exchange league}. In this paper, when a new user connects a BS, the new user will first be assigned to the group with the highest channel gain. So once a user connects/disconnects a BS, there may be new \emph{shift league} or \emph{exchange league}, and the new subchannel assigning vector $\bm{\pi}'$ will not be stable. What we need to do is finding out the \emph{shift leagues} and the \emph{exchange leagues} in $\bm{\pi}'$ and updating the subchannel assigning vector according to these leagues but not re-execute Algorithm 1. To find \emph{i-shift league} or \emph{i-exchange league} in Algorithm $\ref{alg:1}$, graph theory is applied.}

\subsection{Directed Graph}

Let $(n\twoheadrightarrow \tilde{n})$ stand for the operation that moving user $n$ into the group including user $\tilde{n}$, and then moving user $\tilde{n}$ out of its original group.
We assume that $\pi_n=g$, and $\pi_{\tilde{n}}=\tilde{g}$. Then the difference of power consumption of all users on channel $\tilde{g}$ before and after $(n\twoheadrightarrow \tilde{n})$ is
\begin{equation}\small\label{4601}
\setlength\abovedisplayskip{1pt}
\setlength\belowdisplayskip{1pt}
\begin{aligned}
\!\!\bigtriangleup^{\tilde{g}}(n\!\twoheadrightarrow \! \tilde{n},\bm{\pi}_{-\{\!n,\!\tilde{n}\}})\!&=\!\|\overline{\mathbb{A}^{\tilde{g}}}\mathbf{b}^{\tilde{g}}\|_1\Big| (n\!\in\!\mathcal{U}^{\tilde{g},m},\tilde{n}\not\in\mathcal{U}^{\tilde{g},m},\bm{\pi}_{-n})\\
&\!\!-\!\parallel\!\overline{\mathbb{A}^{\tilde{g}}}\mathbf{b}^{\tilde{g}}\parallel_1\!\Big | (n\!\not\in\mathcal{U}^{\tilde{g},m},\tilde{n}\!\in\mathcal{U}^{\tilde{g},m},\bm{\pi}_{-n})
\end{aligned}
\end{equation}

For convenience, users served by BS $m$ are indexed from $1$ to $\|\mathcal{U}^{m}\|$, and $m_k$ denotes the $k$-th user in BS $m$.
{A weighted directed graph $G^m(\mathcal{V},\mathcal{E},\mathcal{W};\bm{\pi})$ is constructed based on  ($ \ref{4601} $), where $\mathcal{V}\!\!=\!\!\mathcal{U}^m$ is the set of nodes, $\mathcal{E}$ denotes the set of edges (these edges only exist between two users served by the same BS and in different groups), $\mathcal{W}$ denotes the set of weights on the edges. The weight on each edge of the weighted directed graph $G^m(\mathcal{V},\mathcal{E};\bm{\pi})$ is calculated according to the adjacent matrix $\mathbb{C}$. The adjacent matrix $\mathbb{C}$ is formed as follows:
\begin{equation}\small\label{4602}
\setlength\abovedisplayskip{1pt}
\setlength\belowdisplayskip{1pt}
\mathbb{C}_{ij}\!=\!\left\{\begin{array}{ll}\!\!\!
\bigtriangleup^{\pi_{m_j}}(m_i\!\twoheadrightarrow\! m_j,\bm{\pi}_{-\{m_i,m_j\}}),
 \!\!\!\!\!~~~~~~& \bm{\pi}_{m_i}\!\neq \!\bm{\pi}_{m_j}\\
\!\!\!\infty,
 & else\\
 \end{array} \right..
\end{equation}
where $\mathbb{C}_{ij}\!=\!\infty$ means that node $m_i$ is not connected with node $m_j$. For instance, if users $m_i$ and $m_j$ are in different groups, the weight on edge $m_i\! -\!\!\!>\! \!m_j$ is $\mathcal{W}_{m_i -> m_j}\!\!=\!\!\mathbb{C}_{ij}\!=\!\bigtriangleup^{\pi_{m_j}}(m_i\twoheadrightarrow m_j,\bm{\pi}_{-\{m_i,m_j\}})$. The adjacent matrix $\mathbb{C}$ is calculated according to the current subchannel assigning vector $\bm{\pi}$. Therefore, although a single node may be involved into many loops, the adjacent matrix $\mathbb{C}$ can be calculated by ($ \ref{4602} $) and the sum of the weight of a loop $m_i \!-\!\!>\!\! m_j-\!\!>\!\!\cdots \!-\!\!\!>\!\! m_k-\!\!\!>\!\! m_j, $ can be calculated by $\mathbb{C}_{ij}\!+\!\cdots\!+\!\mathbb{C}_{ki}$.} Then we have:

\begin{lem1}\label{lem1}
For any $i$ users in different groups of one BS, the following statements are equivalent:

$\textcircled{\small{\small{1}}}$ These $i$ users compose an \emph{i-exchange league}.

$\textcircled{\small{\small{2}}}$ These $i$ users compose a negative loop in graph {$G^m(\mathcal{V},\mathcal{E},\mathcal{W};\bm{\pi})$}.

\end{lem1}

\begin{proof}

We assume that in subchannel assigning vector $\bm{\pi}$, users $n_1,\cdots,n_i$ are in the same BS and in different groups, and we assume $\pi_{n_1}=g_1,\cdots,\pi_{n_i}=g_i$. If we change user grouping vector $\bm{\pi}$ by $\mathbb{J}^{m}\big|\{(n_1\!\!\rightarrow\!\! g_2),\cdots,(n_{i-1}\!\rightarrow \!\! g_i),(n_i\!\!\rightarrow\!\! g_1)\}$, according to ($\ref{4601}$),
$\textcircled{\small{\small{1}}}$ is equivalent to that the difference of total transmit power is negative, i.e.,
{
\begin{equation}\small\label{3005e+6}
\setlength\abovedisplayskip{1pt}
\setlength\belowdisplayskip{1pt}
\begin{aligned}
&\bigtriangleup_{P_{t}}\Big((n_1\rightarrow g_2),\cdots,(n_{i-1}\rightarrow g_i),(n_i\rightarrow g_1)\Big)\\
=&\sum\limits_{k=1}^{i}\bigtriangleup^{g_k}(n_{{k'}}\twoheadrightarrow n_k,\bm{\pi}_{-\{n_{{k'}},n_k\}})<0\\
%=&\sum\limits_{k=1}^{i}\mathcal{C}_{n_{{k'}}}\!({n_{{k'}}}\!\in\!\mathcal{U}^{g_k}\!\!,{n_k}\!\not\in\!\mathcal{U}^{g_k}\!,\bm{\pi}_{\!-\{n_{{k'}},n_k\}}^e\!)\!\\&-\!\sum\limits_{k=1}^{i}\mathcal{C}_{n_k}\!({n_k}\in\mathcal{U}^{g_k}\!,\bm{\pi}_{-{n_k}}^e\!)<0\\
\end{aligned}
\end{equation}}

$\textcircled{\small{\small{2}}}$ is equivalent to that the sum of edge weights of loop $n_1-\!\!>\!n_2-\!\!>\cdots-\!\!>\!n_{i-1}-\!\!>\!n_i-\!\!>\!n_1$ is negative, i.e.,
{
\begin{equation}\small\label{3005e+8}
\setlength\abovedisplayskip{1pt}
\setlength\belowdisplayskip{1pt}
\begin{aligned}
\sum\limits_{k=1}^{i}\mathbb{C}_{{k'}k}=&\sum\limits_{k=1}^{i}\bigtriangleup^{g_k}(n_{{k'}}\twoheadrightarrow n_k,\bm{\pi}_{-\{n_{{k'}},n_k\}})<0\\
%=&\sum\limits_{k=1}^{i}\mathcal{C}_{n_{{k'}}}({n_{{k'}}}\in\mathcal{U}^{g_{{k}}},{n_{k}}\not\in\mathcal{U}^{g_{{k}}},\bm{\pi}_{-\{n_{{k}},n_{{k'}}\}}^e)\\
%&-\sum\limits_{k=1}^{i}\mathcal{C}_{n_{{k'}}}({n_{{k'}}}\in\mathcal{U}^{g_{{k'}}},\bm{\pi}_{-{n_{{k'}}}}^e)<0\\
\end{aligned}
\end{equation}}Therefore, according to ($\ref{3005e+6}$) and ($\ref{3005e+8}$), the proof of Lemma $\ref{lem1}$ is concluded.\qedhere
%According to ($\ref{3005e+7}$),($\ref{3005e+8}$) and ($\ref{3005e+9}$), we have  $\textcircled{\small{\small{1}}}\Leftrightarrow\textcircled{\small{\small{2}}}$.
\end{proof}
We can reduce total transmit power by searching for the \emph{exchange league}s, and changing user grouping strategy according to Definition \ref{defe2}. To find {the} optimal user grouping strategy where number of users in each group is unfixed, \emph{shift league}s {are} needed to be found.

To find the \emph{i-shift league}s by graph theory, we create a virtual user for each group, and let $ \mathcal{N}^v$ denote the set of indexes of virtual users. Then graph {$G^m(\mathcal{V},\mathcal{E},\mathcal{W};\bm{\pi})$} is extended to {$G^m(\mathcal{V}^e,\mathcal{E}^e,\mathcal{W}^e;\bm{\pi}^e)$}, where $\mathcal{V}^e=\mathcal{N}\cup\mathcal{N}^v$, and
$ \bm{\pi}^e $ is the subchannel assigning vector of all users in $\mathcal{V}^e$. %i.e.,$\bm{\pi}^e=\bm{\pi}\cup{\pi_{N+1},\cdots,\pi_{N+G}}$.
%The purpose of adding virtual users are searching for the \emph{shift league}s.

We assume that the target data rates of these virtual users {are} 0. Then according to ($ \ref{3612} $), transmit power of these virtual users is 0, i.e.,
\begin{equation}\small\label{3005e+11}
\setlength\abovedisplayskip{1pt}
\setlength\belowdisplayskip{1pt}
p_{n}=0,~~~~n\in\mathcal{N}^v.
\end{equation}
Therefore, transmit power allocated to the real users will not be affected by the virtual users.

\begin{lem1}
\label{Lemma2}
We can convert any of \emph{shift league}s into \emph{exchange league} with virtual users.
\end{lem1}

\begin{proof}
We assume that in subchannel assigning vector $ \bm{\pi}^e $, users $ n_1, \cdots, n_i $ {are} served by the same BS and can compose an \emph{i-shift league}, i.e., we can reduce total transmit power by operations $(n_1\!\!\rightarrow\!\!\pi_{n_2}),\cdots,(n_{i-1}\!\!\rightarrow \!\!\pi_{n_i})$. We also assume that the virtual user added to the group containing user ${{n_i}}$ is user $n_v$. If we change subchannel assigning vector $\bm{\pi}^e$ by $(n_1\!\rightarrow \!\pi_{n_2}),\cdots,(n_{i-1}\!\rightarrow \!\pi_{n_v}),(n_v\!\rightarrow \!\pi_{n_1})$, since the virtual users will not affect the transmit power of real users, total transmit power will be reduced. In other words, users $n_1, \cdots, n_{i-1}$ can compose an \emph{i-exchange league} with user $n_v$.\qedhere
%\vspace{-0.2cm}
\end{proof}
{
\vspace{-0.3cm}
\setlength{\abovecaptionskip}{-0.02cm}
\setlength{\belowcaptionskip}{-0.1cm}
\begin{figure}[ht]
	\centering
	\includegraphics[scale=0.35]{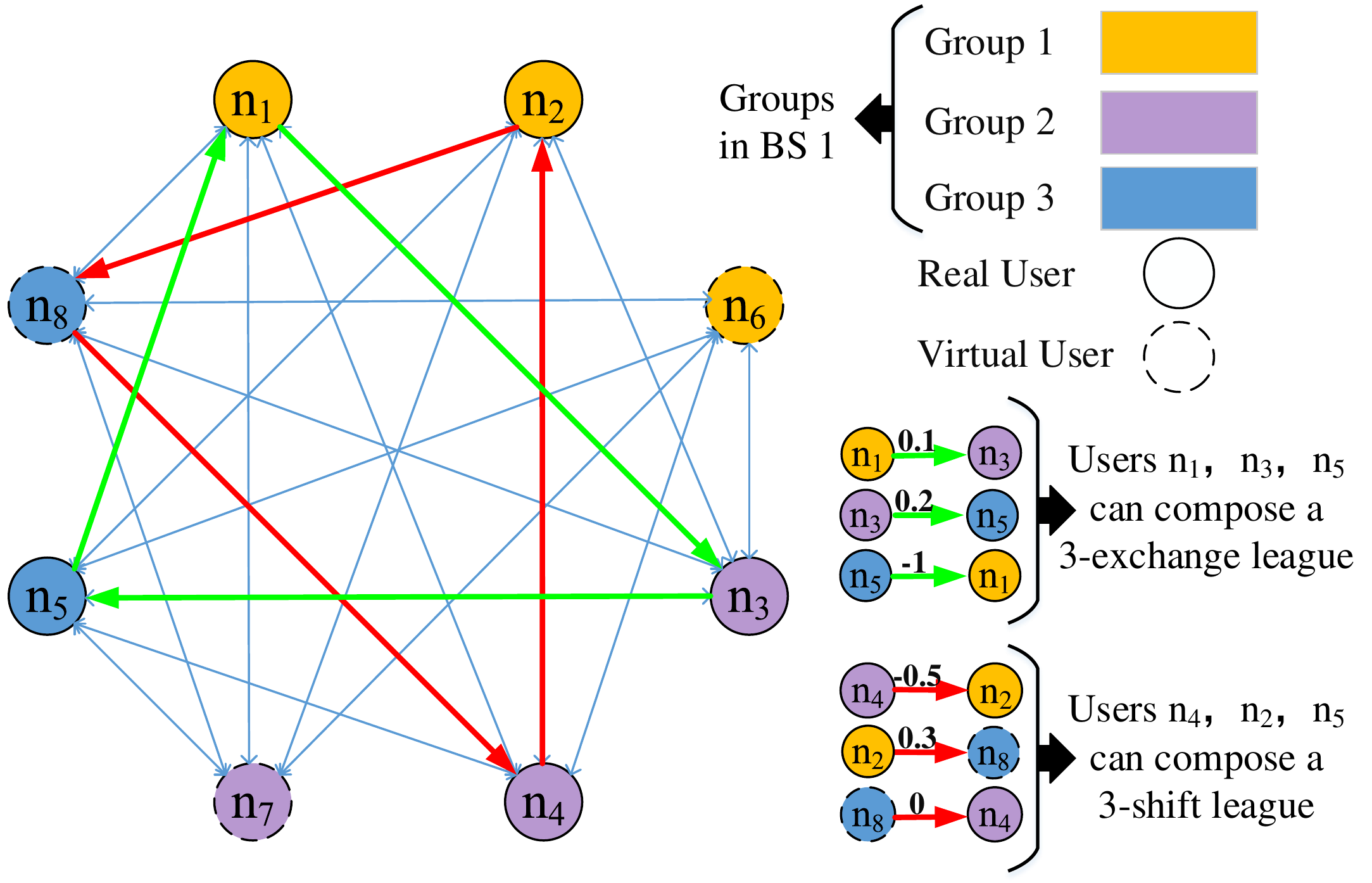}
	\caption{Illustration of Directed Graph}
	\label{fig:7}
\end{figure}}
In order to describe these leagues more intuitively, in Fig. \ref{fig:7}, we show a simple case of the weighted directed graph. In this figure, five users(solid borders) and three virtual users (dashed borders) are assigned into three groups in BS 1, and the edges in this weighted directed graph only exist between the users in different groups. For convenience, the negative loop composed by users in the same BS and in different groups is called \emph{negative differ-group loop}. In this figure, $n_1-\!\!>\!n_3\!-\!\!>\!n_5-\!\!>\!n_1$ is a \emph{negative differ-group loop}, and all these three users are real users, so these three users can compose a 3-\emph{exchange league}. Therefore, total transmit power can be reduced by operations $(n_1\!\rightarrow \!\pi_{n_3}),(n_{3}\!\rightarrow \!\pi_{n_5}),(n_5\rightarrow \pi_{n_1})$. In addition, $n_4-\!\!>\!n_2\!-\!\!>\!n_8-\!\!>\!n_4$ is a \emph{negative differ-group loop}, where user $n_8$ is a virtual user in group 3, so users $n_4,n_2,n_5(n_8)$ compose a 3-\emph{shift league} ($n_8,n_5\in \mathcal{U}^3$). That means total transmit power can be reduced by operations $(n_4\!\!\rightarrow
\!\pi_{n_2}),(n_{2}\!\!\rightarrow \!\!\pi_{n_8})$. Combine Lemma 1 and Lemma 2 with Definition \ref{defe1}-\ref{defe3}, we can obtain Theorem $\ref{thm+1}$ and Theorem $\ref{thm+111}$.

\begin{thm1}\label{thm+1}
In weighted directed graph {$G^m(\mathcal{V}^e,\mathcal{E}^e,\mathcal{W}^e;\bm{\pi}^e)$}, if no \emph{negative differ-group loop} can be found,
the user grouping strategy based on subchannel assigning vector $\bm{\pi}^e$ is an \emph{all-stable solution}.
%If there is no \emph{negative differ-group loop} in the directed graph {$G^m(\mathcal{V}^e,\mathcal{E}^e,\mathcal{W}^e;\bm{\pi}^e)$}, the grouping strategy of real users $\bm{\pi}$ is
\end{thm1}

\begin{proof}
For any $ i $ users that can compose an \emph{i-exchange league} in the user grouping strategy based on subchannel assigning vector $\bm{\pi}$, on the basis of Lemma 1, these users can compose a \emph{negative differ-group loop} in weighted directed graph {$G^m(\mathcal{V}^e,\mathcal{E}^e,\mathcal{W}^e;\bm{\pi}^e)$}. Similarly, For any $ i $ users that can compose an \emph{i-shift league} in the user grouping strategy based on subchannel assigning vector $\bm{\pi}$, on the basis of Lemma 1 and Lemma 2, these users can compose a \emph{negative differ-group loop} in weighted directed graph {$G^m(\mathcal{V}^e,\mathcal{E}^e,\mathcal{W}^e;\bm{\pi}^e)$}. Therefore, if no \emph{negative differ-group loop} can be found in graph {$G^m(\mathcal{V}^e,\mathcal{E}^e,\mathcal{W}^e;\bm{\pi}^e)$}, no \emph{i-shift league} or \emph{i-exchange league} can be found in the user grouping strategy based on subchannel assigning vector $\bm{\pi}^e$. So the proof of Theorem $ \ref{thm+1} $ is concluded. \qedhere
\end{proof}

\begin{thm1}\label{thm+111}
In weighted directed graph {$G^m(\mathcal{V}^e,\mathcal{E}^e,\mathcal{W}^e;\bm{\pi}^e)$}, if users $ n_1, \cdots, n_i $ can compose a \emph{negative differ-group loop} $n_1\!-\!>\!\cdots\!-\!\!>\!n_i-\!\!>\!n_1$, total transmit power can be reduced by operations $(n_1\!\rightarrow\! \pi_{n_2}),\cdots,(n_{i-1}\!\rightarrow\! \pi_{n_i}),(n_i\!\rightarrow\! \pi_{n_1})$.
\end{thm1}

\begin{proof}

In weighted directed graph {$G^m(\mathcal{V}^e,\mathcal{E}^e,\mathcal{W}^e;\bm{\pi}^e)$}, if users $ n_1, \cdots, n_i $ can compose a \emph{negative differ-group loop} $n_1\!-\!>\!\cdots\!-\!\!>\!n_i-\!\!>\!n_1$, combine ($\ref{3005e+6}$) with ($\ref{3005e+8}$), we have
\begin{equation*}\small\label{3005e+21}
\begin{aligned}
\setlength\abovedisplayskip{1pt}
\setlength\belowdisplayskip{1pt}
\sum\limits_{k=1}^{i}\!\mathbb{C}_{{k'}k}
\!=\!\bigtriangleup_{P_{t}}\!\Big(\!\!(n_1\!\!\rightarrow\!\! \pi_{n_2}\!),\!\cdots\!,\!(n_{i-1}\!\!\rightarrow\!\! \pi_{n_i}\!),(n_i\!\!\rightarrow\!\! \pi_{n_1}\!)\!\!\Big)\!\!<\!\!0.
\end{aligned}
\end{equation*}
In other words, we can reduce total transmit power allocated to all real users and virtual users by operations $ (n_1\rightarrow \pi_{n_2}),\cdots,(n_{i-1}\rightarrow \pi_{n_i}),(n_i\rightarrow \pi_{n_1})$. In addition, according to ($ \ref{3005e+11}$), transmit power allocated to the virtual users is 0. Therefore, we can reduce total transmit power allocated to all real users by operations $ (n_1\rightarrow \pi_{n_2}),\cdots,(n_{i-1}\rightarrow \pi_{n_i}),(n_i\rightarrow \pi_{n_1})$. So the proof of Theorem $ \ref{thm+111} $ is concluded. \qedhere
\end{proof}

\subsection{Algorithm Description}\label{section5B}

\begin{algorithm}[htbp]
 \caption{Extended BellmanFord Algorithm(EBA) for Finding the \emph{Negative Differ-Group Loop}}\label{alg:2}
  \KwIn{Graph {$G^m(\mathcal{V}^e,\mathcal{E}^e,\mathcal{W}^e;\bm{\pi}^e)$},

  \quad \quad \quad Adjacent matrix$\mathbb{C}=(\mathbb{C}_{ij})_{i,j\in\mathcal{V}^e}$}
  \KwOut{\emph{Negative differ-group loop} $\mathcal{L}$}

  %Let the distance from super node to node $n$ be $\!m_n\!\!=\!\!0$, the path from super node to node $n$ be $\mathcal{T}_n=\varnothing$, $n\in \mathcal{V}$;
$\!c_n\!\!=\!\!0$, $\mathcal{T}_n=\varnothing$, $n\in \mathcal{V}$;

  \Repeat{$\mathcal{T}_n$, $ n\in\mathcal{V}$ do not change}{
        \For{$i=1 \textrm{ to }\|\mathcal{V}^e\|$}{
        %\IF{$c_{n}>c_{\tilde{n}}+\mathbb{C}_{\tilde{n}n}$}
        \For{$j=1 \textrm{ to }\|\mathcal{V}\|$}{
        \If{$(c_{j}\!>\!c_{i}\!+\!\mathbb{C}_{ij}) \&  (\mathbb{C}_{kj}\!\neq \!\infty, \forall k\in\mathcal{T}_i\setminus \{j\})$}{
         $\mathcal{T}_j=\mathcal{T}_i\cup \{i\}$, $c_{j}=c_{i}+\mathbb{C}_{ij}$
        }
        \If{$(c_{j}\!>\!c_{i}+\!\mathbb{C}_{ij}) \& (\exists \mathbb{C}_{kj}\!= \!\infty,k\in\mathcal{T}_i\!\setminus \!\{j\})$}{
         Find the shortest path $\mathcal{T}'_i$ from the super node to user $i$ with $ \mathbb{C}_{kj}\neq \infty, \forall k\in\mathcal{T}'_i\setminus \{j\}$, assume the distance of path $\mathcal{T}'_i$ is $m'_{i}$;

        \If{$c_{j}>c'_{i}+\mathbb{C}_{ij}$}{
         $\mathcal{T}_j=\mathcal{T}'_i\cup \{i\}$, $c_{j}= c'_{i}+\mathbb{C}_{ij}$
		}
        }
        \If{$\|\mathcal{T}_j\|> G$}{
         Find the \emph{negative differ-group loop} $\mathcal{L}$ in $\mathcal{T}_j$;
         \textbf{return} $\mathcal{L}$ and break;
		}

		}

		}
          }

\end{algorithm}

As stated in Theorem $\ref{thm+1}$ and Theorem $\ref{thm+111}$, the action $\mathbb{J}^{m}\big|_{\mathcal{D}_m(\bm{\pi}\xrightarrow[]{\mathbb{J}^{m}}\bm{\pi}')<0}$ in Algorithm $\ref{alg:1}$ can be found by searching for the \emph{negative differ-group loop}s in weighted directed graph
{$G^m(\mathcal{V}^e,\mathcal{E}^e,\mathcal{W}^e;\bm{\pi}^e)$} and update user grouping strategy according to these loops, until no \emph{negative differ-group loop} can be found in the graph. In order to find these loops, extended Bellman-Ford algorithm (EBA) is described in Algorithm $\ref{alg:2}$ in detail\cite{fakcharoenphol2006planar}.
We first add a super node to this graph, the super node has outgoing edges to all the other nodes. The main procedure of EBA is searching for the shortest path from the super node to all other nodes and relaxing the outgoing edges from the super node until no outgoing edges can be relaxed\cite{bellman1958routing}. When relaxing these outgoing edges, users in the same group are avoided to appear in the same path, which is the main difference between EBA and original Bellman-Ford algorithm. Then if the length of the shortest path from the super node to a node is greater than $G$, in this path, there must exist
a \emph{negative differ-group loop} in this graph. In EBA, we relax the shortest path from the super node to node $n$, i.e., $\mathcal{T}_n$, ($\forall n\in \mathcal{V}^e$) in step 6 or step 11 repeatedly until $\mathcal{T}_n$ ($\forall n\in \mathcal{V}^e$) does not change any more. Step 9 is a recursive procedure by calling EBA repeatedly. Hence the computational complexity of EBA increases exponentially with the dimensions of $G$ and the number of users in each group.

%\vspace{-0.3cm}

%\vspace{-0.3cm}
\begin{algorithm}[htbp]
 \caption{Fast Greedy-Strategy(FGA) for Finding the \emph{Negative Differ-Group Loop}}\label{alg:FGA}
  \KwIn{Group $ G(\mathcal{V},\mathcal{E},\mathcal{W};\bm{\pi}^e)$,

  \quad \quad \quad Adjacent matrix$\mathbb{C}=(\mathbb{C}_{ij})_{i,j\in\mathcal{V}}$}
  \KwOut{\emph{Negative differ-group loop} $\mathcal{L}$}
  %\textbf{initialization:} Let the iteration index $k=1$.
  $\mathcal{T}=\varnothing$,$c_t= 0$;

 $c_s= 0$;
        %\STATE Set $\mathcal{M}=\mathcal{V}$,$\mathbf{T}=(t_g)_{g\in \mathcal{G}}=(0)_{1\times G}$,$\delta=\delta_{min}=0$;%,$\mathcal{A}=\{\mathbb{C}_{ij}^e\}_{i,j\in\mathcal{V}^e}$

        \For{$x=1 \textrm{ to }\lceil\alpha(N+G)\rceil$}{
        $\mathcal{T}=\varnothing$;

         Find the minimal edge $\mathbb{C}_{ij}=\min\limits\{\mathbb{C}_{ij}|i,j\in\mathcal{V}\}$;

        $\mathcal{T}=\mathcal{T}\cup \{i\}\cup \{j\}$, $c_t= \mathbb{C}_{ij}+\mathbb{C}_{ji}$;

        $\mathbb{C}_{ij}=\infty$, $i= j$;
        %\STATE Find the minimal edge $\mathbb{C}_{ij}=\min\limits\{\mathbb{C}_{ij}|i,j\in\mathcal{V}\}$;

        \For{$l=3 \textrm{ to }G$}{
        Find the minimal output edge of node $i$ : $\mathbb{C}_{ij}=\min\limits\{\mathbb{C}_{ij}|j\in\mathcal{V}, \mathbb{C}_{jk}\neq\infty, \forall k\in \mathcal{T}\}$;

        $c_t= c_t+\mathbb{C}_{ij}$, $\mathcal{T}=\mathcal{T}\cup \{j\}$;

        \If{$c_s>c_t+\mathbb{C}_{ji}$}{
        $\mathcal{L}=\mathcal{T}$, $c_s= c_t+\mathbb{C}_{ji}$;
		}
        $i= j$;
        }
        }
        %\IF{$\delta_{min}<0$}
        %\STATE Add the first recorded node in $\mathcal{L}$ to $\mathcal{Q}$;
        \textbf{return} $\mathcal{L}$.

\end{algorithm}

In order to reduce the computational complexity of EBA, we also design a suboptimal fast strategy based on a greedy strategy (FGA) to search for these \emph{negative differ-group loop}s as described in Algorithm $\ref{alg:FGA}$.
%To approach the solution of the problem in ($\ref{3619}$) with polynomial-time computational complexity, we design a suboptimal fast strategy based on a greedy strategy to find the \emph{negative differ-group loop}s.
In FGA, we first initialize some variables in step 1-2, where $\mathcal{T}$ records the searching path in each iteration, $c_t$ is the sum weights of edges in this path, and $c_s$ is the sum weights of edges in the loop $\mathcal{L}$ which is the minimal loop we have found. Second we look for the minimum edge $\mathbb{C}_{n_1n_2}$ in $\mathbb{C}$, and calculate the sum weights of edges in the loop composed by user $n_1$ and $n_2$, i.e., $n_1-\!\!>\!n_2-\!\!>\!n_1$, in step 5. This loop is recorded in $\mathcal{T}$ as step 6 stated. In step 9-14, we look for the minimum output edge $\mathbb{C}_{n_2n_3}$($\pi_{n_3}\neq\pi_{n_1},\pi_{n_3}\neq\pi_{n_2}$) of node $n_2$, and calculate the sum weights of edge in loop
$n_1-\!\!>\!n_2-\!\!>\!n_3-\!\!>\!n_1$. The operations of looking for next minimum output edge and calculating sum weights are repeated until there is no next output edge. If the loop $\mathcal{T}$ in step 10 is negative, we can output this loop as \emph{negative differ-group loop}. Steps $4-15$ are repeated $\lceil\alpha(\|\mathcal{U}^m\|+G)\rceil$ times ($\alpha\in [\tfrac{1}{\|\mathcal{U}^m\|+G},\|\mathcal{U}^m\|]$), {where $\alpha$ is a regulating factor to reduce unnecessary computation. As shown in the simulation results in \ref{sction6A}, the total transmit power will not be reduced with the number of iterations  after only several iterations. Therefore, unnecessary computation can be reduced with very few transmit power loss by adjusting regulating factor $\alpha$.}

{\emph{Computational complexity Comparison of FGA and the EBA:} In this paper, step 9 of Algorithm 2 (EBA) is a recursive procedure by calling EBA repeatedly. Hence the computational complexity of EBA increases exponentially with the dimensions of $G$ and the number of users in each group. In order to reduce the computational complexity of EBA, we also design a suboptimal fast algorithm based on a greedy strategy (FGA) to search for these \emph{negative differ-group loop}s as described in Algorithm 3. In FGA, the computational complexity of steps $ 8 $-$ 15 $ is $O\big(G(G+N)\big)$. Therefore, the computational complexity of FGA is $O\big(G(G+N)^2\big)$. If FGA repeats $C$ times in Algorithm 1, the computational complexity of the proposed fast user grouping strategy is $O\big(CGM(G+N)\big)$. According to the simulation results in section VI, $C$ is much less than $(G+N)$.}

\begin{figure*}[htb]
\setlength{\belowcaptionskip}{-0.5cm}
\setlength{\belowdisplayskip}{-1.2cm}
%\hspace{0.7in}
%\begin{minipage}{0.3\linewidth}
\centering
\subfigure[]{
\label{fig:dt11}
\includegraphics[height=0.25\textwidth]{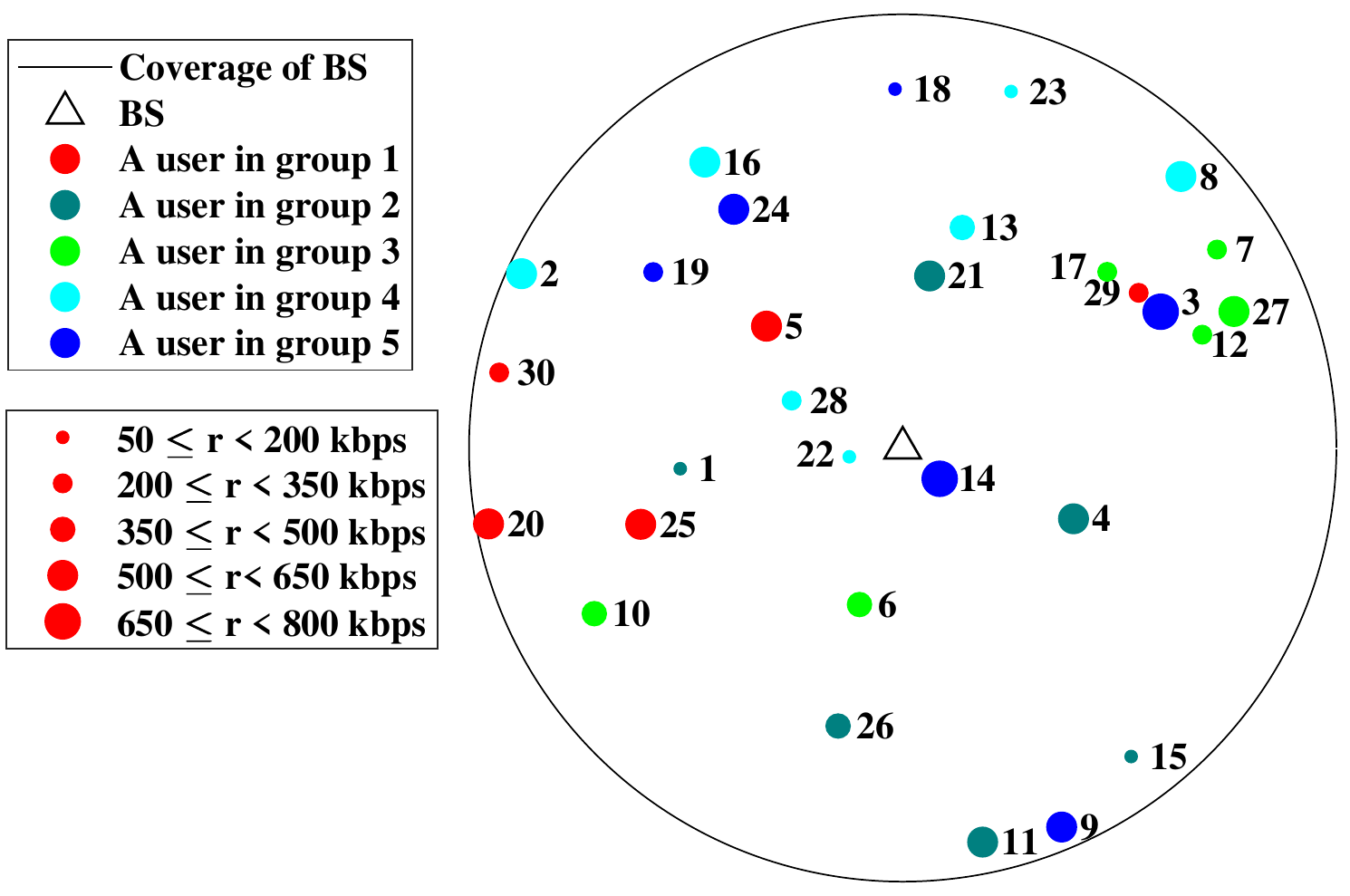}}
\subfigure[]{
\label{fig:dt211}
\includegraphics[height=0.25\textwidth]{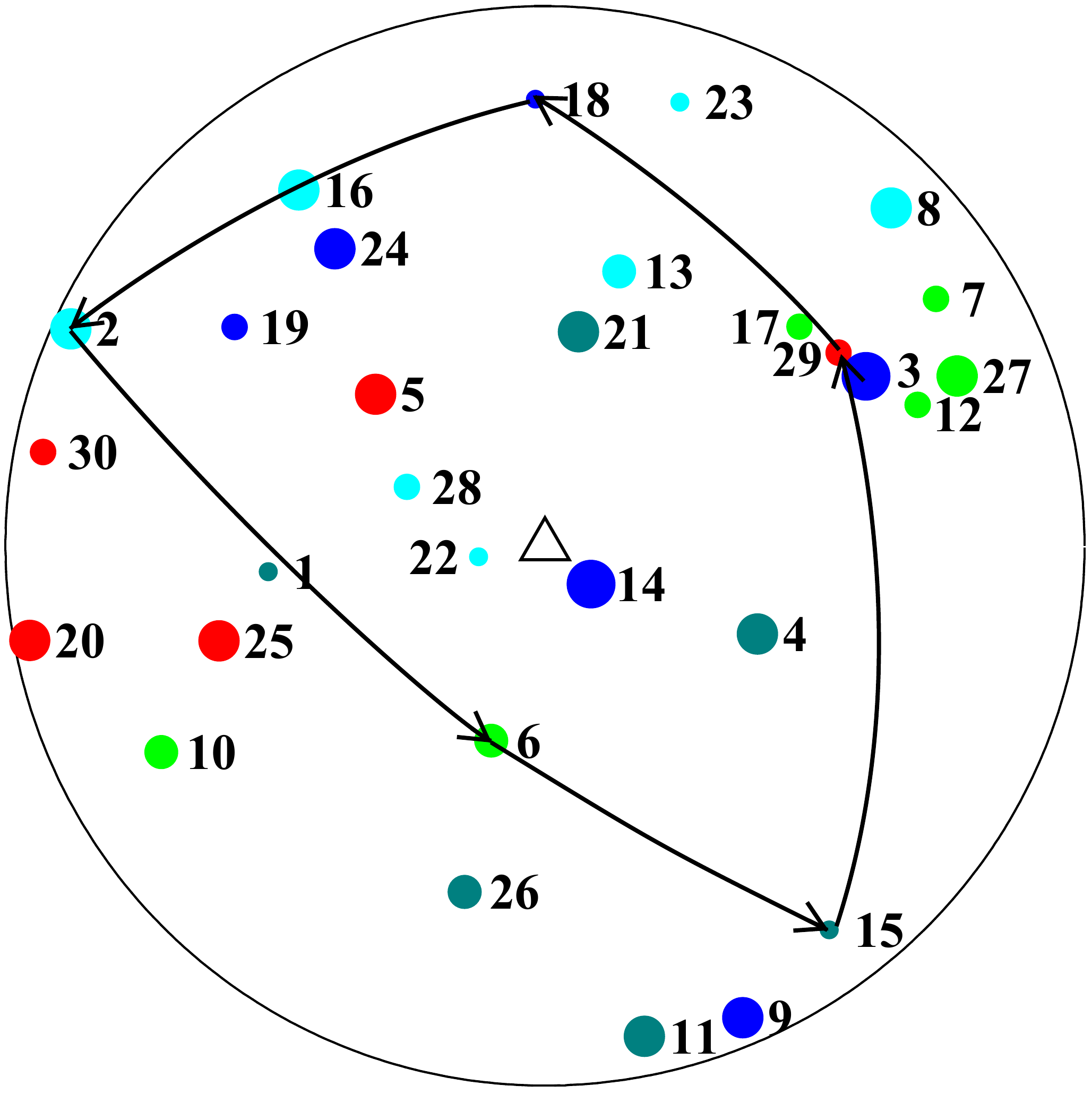}}
\subfigure[]{
\label{fig:dt311}
\includegraphics[height=0.25\textwidth]{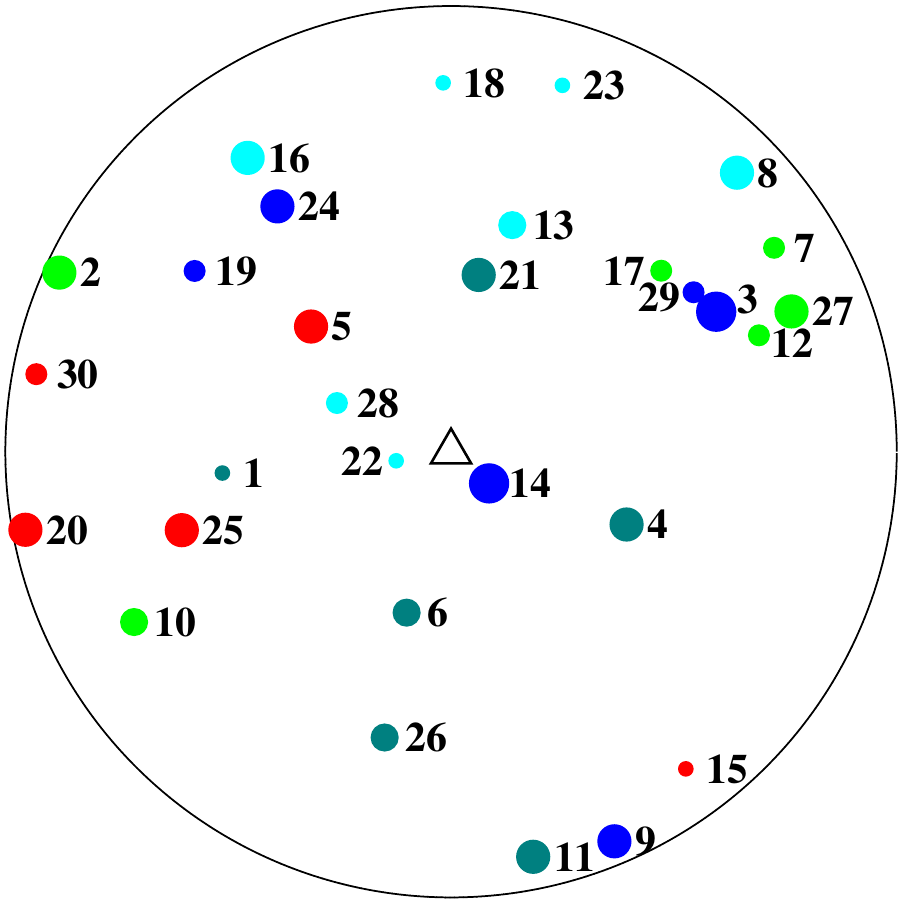}}
%\caption{Detailed Procedure of EBA Strategy(Updated)}
\centering
\caption{Detailed Procedure of EBA Strategy: a. Initialization; b. Negative Differ-group Loop:$2-\!\!>\!6-\!\!>\!15-\!\!>\!29-\!\!>\!18-\!\!>\!2$; c. Updated}.
\label{fig:dt00}
%\end{minipage}%
\end{figure*}

%(150m, 450m), (450m, 450m), (150m, 150m), and (450m, 150m)

%

%\subsection{A Simple Example of Reducing Total Transmit Power}\label{section5C}
\subsection{The Relationship between EBA and FGA with An Example}\label{section5C}

{ EBA and FGA are two algorithms to find an appropriate action for each player in the potential game, i.e., to find the user grouping strategies of users associated to each BS. Based on the potential game model, the grouping strategy of each BS is chosen one by one according to the strategies of the other BSs. In this paper, the problem of finding an appropriate user grouping strategy for each BS is converted into the problem of searching for negative differ-group loops in the graph composed of users. This problem is solved by EBA and FGA.
%It is worth noting that both EBA and FGA algorithms can be used in single-cell scenario to group users as stated in our preliminary work in \cite{guo2019interference}.
We show the detailed procedure of reducing total transmit power by EBA algorithm in Fig. \ref{fig:dt00} in single-cell scenario. The simulation settings are consistent with section \ref{section6}.
As shown in Fig. \ref{fig:dt00}, when a negative differ-group loop $2$(group 4)$-\!\!>\!6$(group 3)$-\!\!>\!15$(group 2)$-\!\!>\!29$(group 1)$-\!\!>\!18$(group 5)$-\!\!>\!2$(group 4) is found as shown in Fig. \label{fig:dt211}, if we change the user grouping strategy by putting user $2$ into group $3$, putting user $6$ into group $2$, putting user $15$ into group $1$, putting user $29$ into group $5$, and putting user $18$ into group $4$ (the updated grouping strategy is shown in Fig. \ref{fig:dt311}), the total transmit power will be reduced. In order to find those negative differ-group loops, the EBA algorithm is first proposed.

The main procedure of EBA is searching for the shortest path from the super node to all other nodes, i.e., all users, and avoiding the users in the same group being putted into this path. If there is a negative loop with user $n$ in this graph, the shortest path from the super node to user $n$ will be negative and the number of nodes of this path will be larger than the number of groups. The reason is that, this shortest path can be reduced by crossing this negative differ-group loop many times. Therefore, we can find the negative differ-group loop by searching for the shortest path from the super node to all other nodes until this path is longer than the number of groups.

The EBA algorithm can find all the negative differ-group loop in this graph, however,  the computational complexity of EBA increases exponentially with the dimensions of G and the number of users in each group. In order to reduce the computational complexity of EBA, we also design a suboptimal fast strategy based on a greedy strategy (FGA) to search for these negative differ-group loops. The main procedure of FGA is searching for the minimal edge (e.g. $i-\!\!>\!j$) in this graph firstly, and then looking for the node $k$ with the minimal edge to node $j$ as the next hop node. The latter is  repeated until there is no next output edge or the negative differ-group loop that can be  found.}

\section{Performance Evaluation}\label{section6}

In this section, the performance of the proposed two user grouping strategies, i.e., EBA and FGA, in downlink multi-cell NOMA systems is evaluated by simulations. In the simulations, four BSs are respectively placed in (250m, 250m), (750m, 250m), (250m, 750m), and (750m, 750m) and some users are randomly distributed in a 1000m$\times$1000m rectangular area around the BSs. We set the minimum distance from user to BS to 15m.
The large-scale channel gain is $128.1\!+\!37.6$log$_{10}( d_{n}[$km$])$ dB.
The Rayleigh fading coefficient follows an i.i.d. Gaussian distribution as $\beta\!\!\thicksim\!\mathcal{CN}(0, 1)$. We set the noise power to $\sigma^2=BN_0$, where $B\!\!= \!\!200$ kHz and $N_0\!= $-$174$ dBm/Hz. The targeted data rate of each user is randomly chosen from 60 to 600 kbps. %The performance of two proposed strategies, i.e., EBA and FGA, is evaluated by simulations.

\subsection{Convergence Performance}\label{sction6A}

\begin{figure*}[htb]
\setlength{\belowcaptionskip}{-0.5cm}
\setlength{\belowdisplayskip}{-1.2cm}
%\hspace{0.7in}
%\begin{minipage}{0.3\linewidth}
\centering
\subfigure[]{
\label{fig:singledd}
\includegraphics[height=0.26\textwidth]{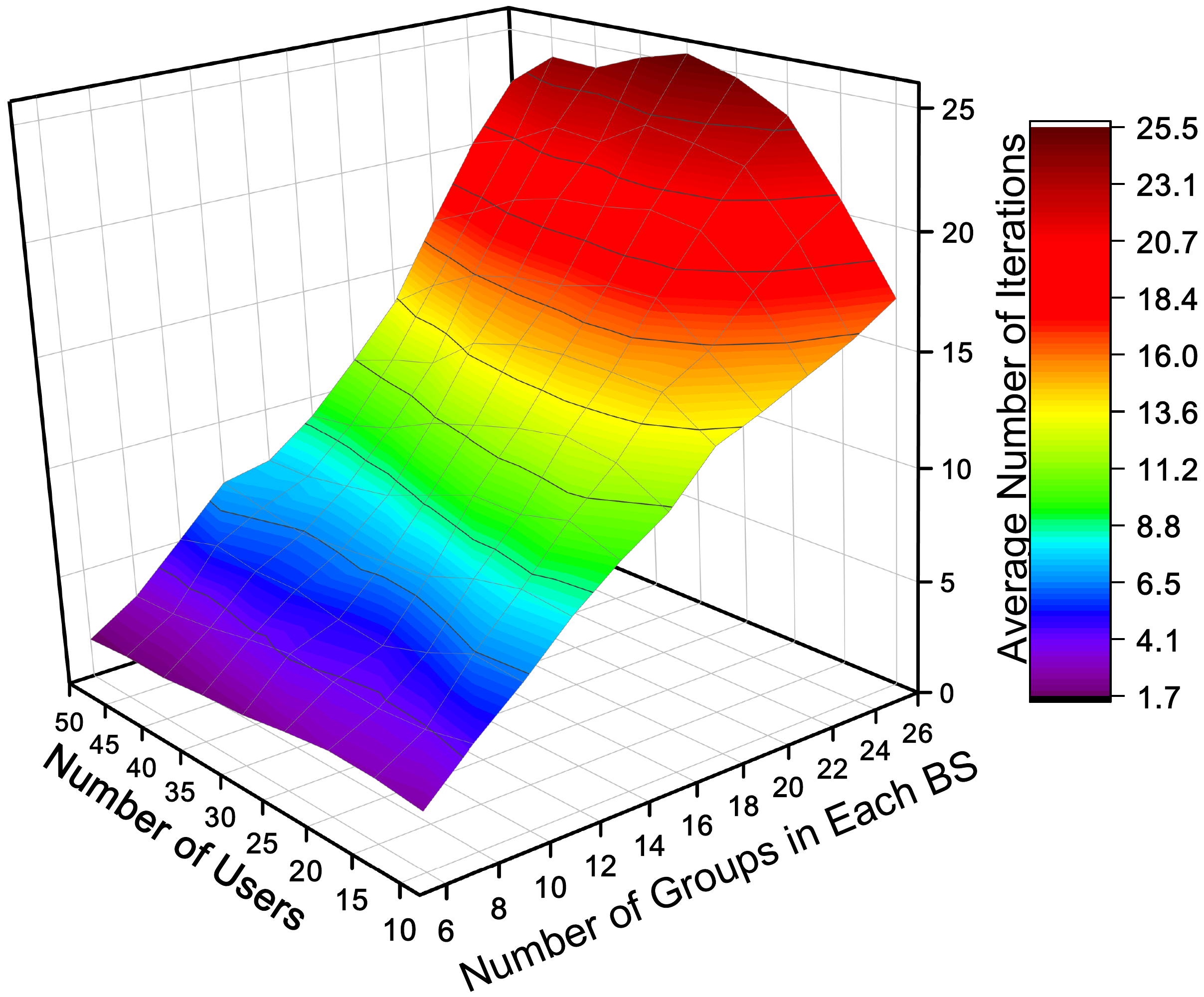}}
\subfigure[]{
\label{fig:3}
\includegraphics[height=0.26\textwidth]{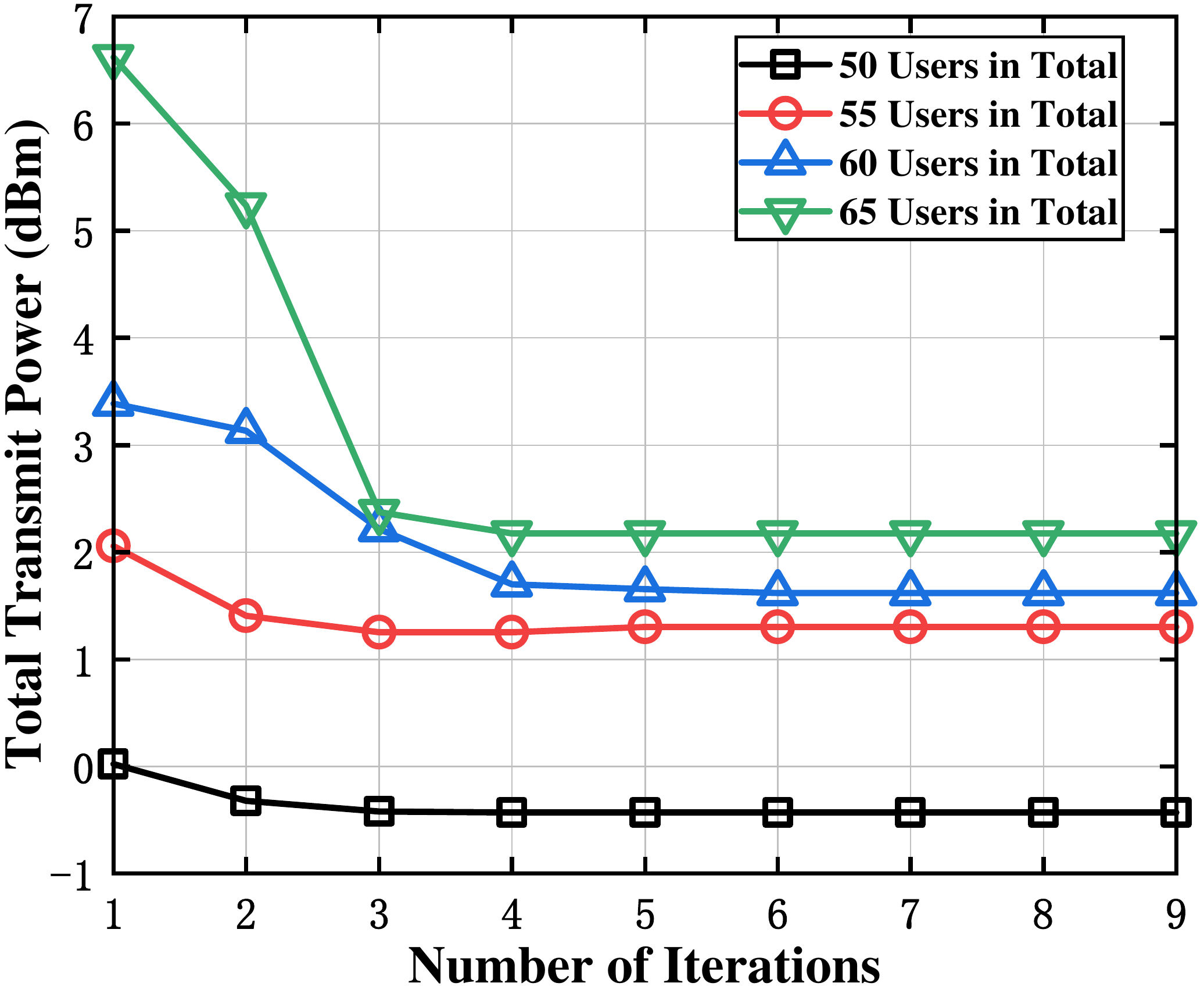}}
%\caption{Detailed Procedure of EBA Strategy(Updated)}
\centering
\caption{Convergence Performance of FGA: a. Average Number of Iterations (Single-Cell); b. Total Transmit Power vs. Number of Iterations (Multi-Cell).}
\label{fig:shoulian}
%\end{minipage}%
\end{figure*}

We first compute the average number of iterations of FGA to stop in single-cell scenario. In the simulations of single-cell scenario, the BS is located in the cell center and the users are randomly distributed in a circular range with a radius of 500m. We vary the number of groups and the number of users. The results presented in Fig. \ref{fig:singledd} show that the average number of iterations slowly increases when the number of groups increases. This is because that the size of adjacent matrix will grow as the number of groups increases. In addition, the average number of iterations of our strategy is below 26 in the scenario where 26 groups and 50 users in total exist. The reason is that, in each iteration of FGA, we try to search for the minimum negative differ-group loop to accelerate convergence. Fig. \ref{fig:3} shows total transmit power for all users with iterations when grouping users by FGA in multi-cell scenario, where $ G=10 $, $ N=[50 ,\ldots , 65] $. We can see that total transmit power declines rapidly and tends to a stable value after only several iterations. These results verify the statement that $C$ is much less than $(G + N )$ in section \ref{section5B}.

\subsection{Impacts of $ \alpha $}\label{sction6B}

\begin{figure*}[htb]
\vspace{-0.6cm}
\setlength{\belowcaptionskip}{-0.5cm}
\setlength{\belowdisplayskip}{-1.2cm}
%\hspace{0.7in}
%\begin{minipage}{0.3\linewidth}
\centering
\subfigure[]{
\label{fig:alfa_single}
\includegraphics[height=0.26\textwidth]{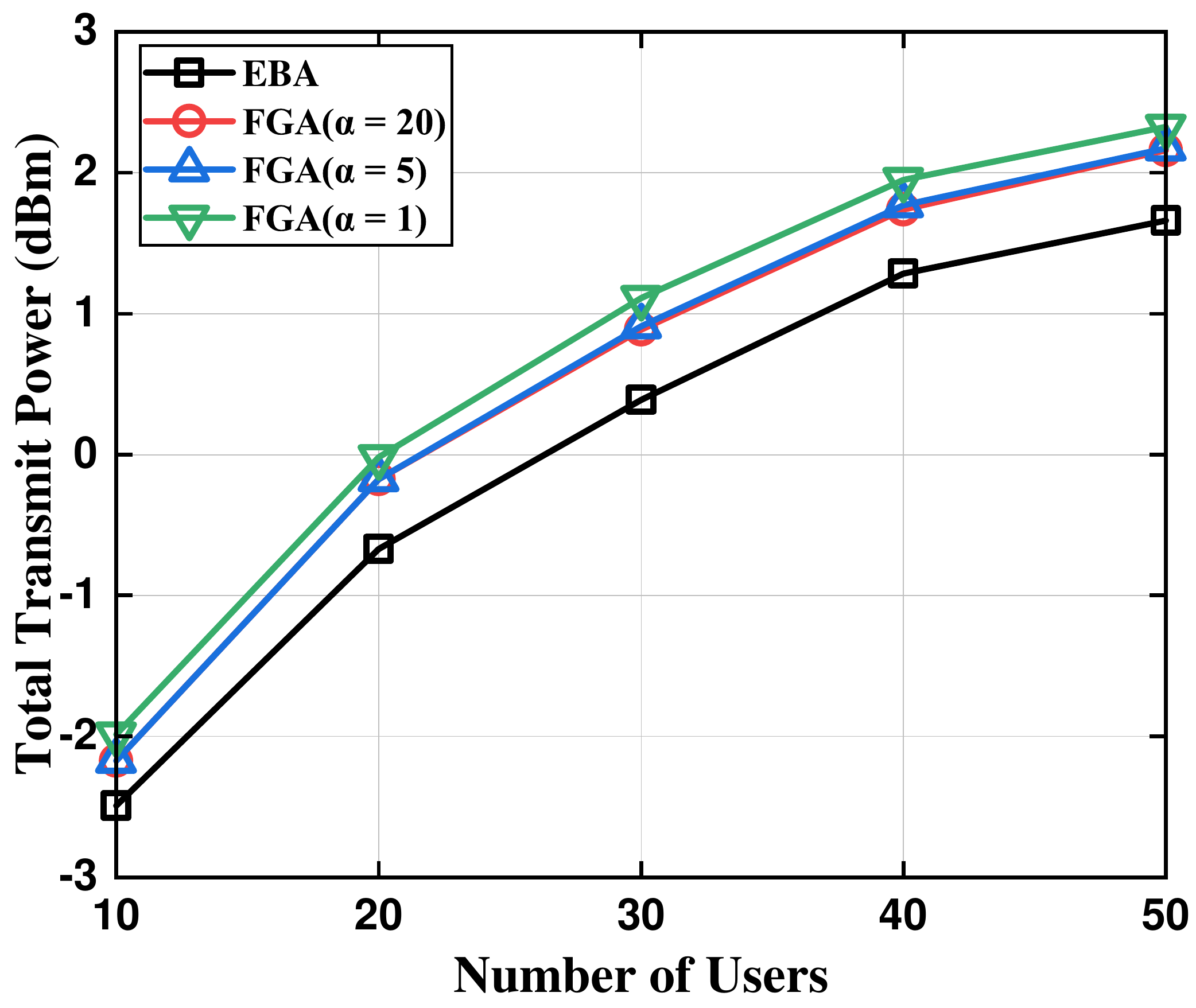}}
\subfigure[]{
\label{fig:6}
\includegraphics[height=0.26\textwidth]{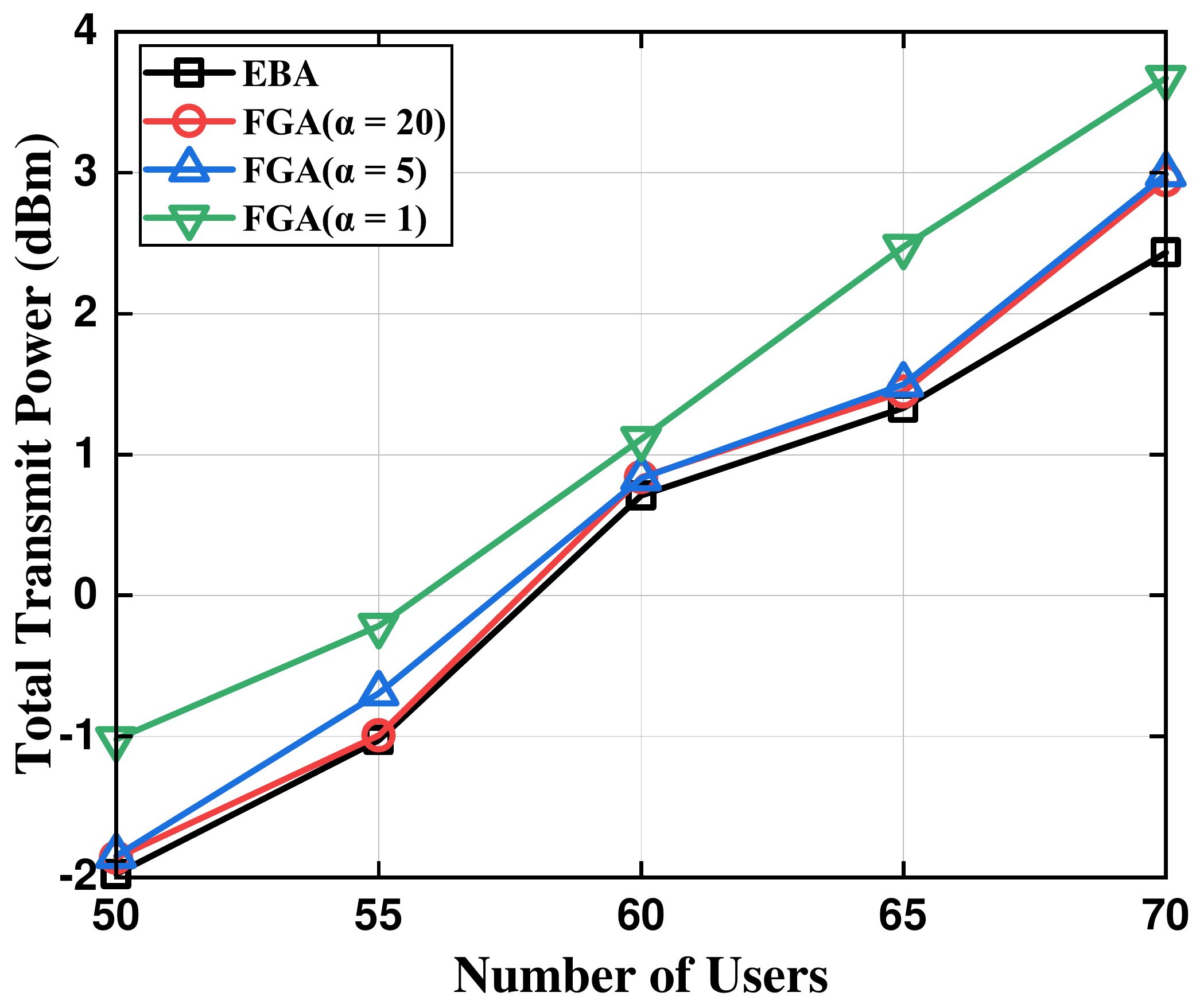}}
%\caption{Detailed Procedure of EBA Strategy(Updated)}
\centering
\caption{Impacts of $ \alpha $ on FGA: a. Total Transmit Power vs. Number of Groups (Single-Cell); b. Total Transmit Power vs. Number of Users (Multi-Cell).}
\label{fig:alfaaaaa}
%\end{minipage}%
\end{figure*}

In FGA strategy, $\alpha$ is a regulating factor to limit the number of iterations. If $\alpha$ increases, more negative differ-group loops might be found and computational complexity will increase, then more transmit power might be reduced. In Fig. \ref{fig:alfa_single}, we show total transmit power in single-cell scenario with $G=[10, 20, \cdots, 50]$ and number of users being two times of groups. The results show that total transmit power increases with $G$, decreases slowly with the increase of $\alpha$ and almost changes no more when $\alpha\geq 5$. Fig. \ref{fig:6} shows total transmit power in multi-cell scenario with $N=[50, 55, \cdots, 70]$ and $G=10$. In this figure, power consumption also decreases slowly with the increase of $\alpha$ and almost changes no more when $\alpha\geq 5$. These results show that impacts of $ \alpha $ will not change as the increase of the number of BSs. Note that performance of FGA($\alpha=5$) approaches that of EBA. Therefore in the following simulations, we set $\alpha=5$. {Furthermore, we can see that compared with OMA system, more users can be served at the same resource block in NOMA system (only one user can be served at the same resource block in OMA system).}

\subsection{Performance Comparison}

\begin{figure}[ht]\nonumber
\vspace{-0.6cm}
\setlength\abovedisplayskip{1pt}
\setlength\belowdisplayskip{1pt}
	\centering
	\includegraphics[height=0.26\textwidth]{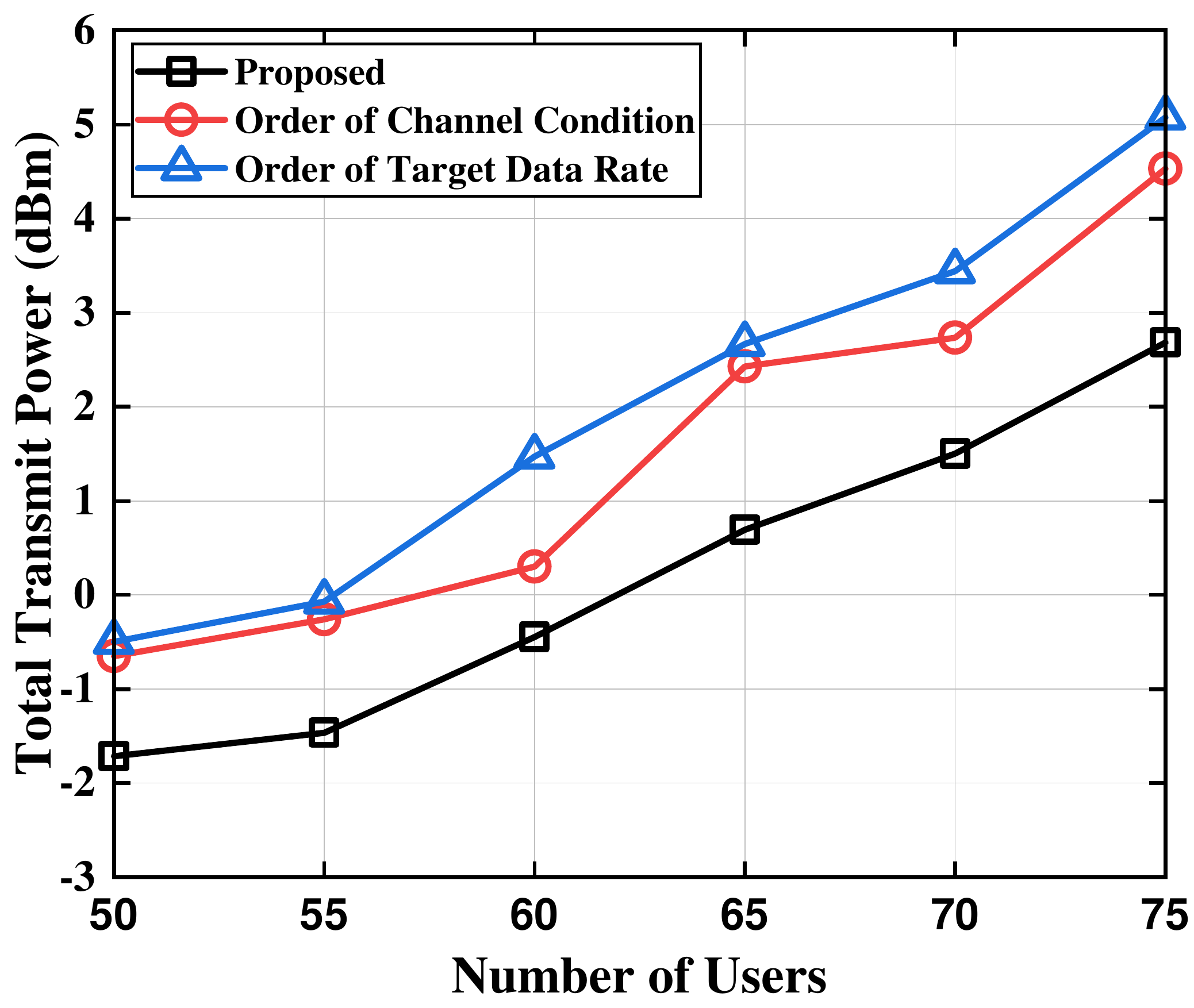}
	\caption{Impacts of Decoding Order. (Total Transmit Power vs. Number of Users)}
	\label{fig:16}
\end{figure}

{As stated in section III, due to severe inter-cell interference, the traditional SIC decoding order based on channel condition is no longer applicable. We compare the performance of the proposed SIC decoding order stated in Theorem 1 with two different SIC decoding orders, one is the order of channel coefficient, and the other one is the descending order of target data rate. The logic behind this is that, in general, the user with high target data rate need more transmit power from BS and will bring more interference to the users being decoded after this user. To reduce those interference, the signals of the users with higher target data rates are set to be decoded firstly. We varied the number of users from 50 to 75. Fig. \ref{fig:16} shows total transmit power of all users for the three orders. We can see that the proposed decoding order outperforms the other two decoding orders.
It is noted that we should allocate power to users according to their SIC decoding order to make sure every user can correctly detect its information. Different power allocation orders lead to different total transmit power and it is important to take inter-cell interference into consideration when allocating power. Therefore, in order to minimize total transmit power, it is essential to consider the impact of inter-cell interference when choosing SIC decoding order. This is the reason that the proposed decoding order outperforms the other two decoding orders.}

Next, the proposed two strategies are compared with two existing user grouping strategies, where one is selecting users whose channel conditions are more distinctive into the same group (``SCCD'') \cite{Zhang2017k,Ali1}, {with SCCD, the user with the highest channel gain is grouped with the user with the lowest channel gain, and the user with the second highest channel gain is grouped with the user with the second lowest channel gain, and so on; the other one is based on Gale-Shapley algorithm(``Gale-Shapley'') \cite{zhang1}. With Gale-Shapley algorithm, users and groups are considered as two player sets with their own preference lists. Each user prefers the group with better channel condition and each group prefers the user with higher channel gain. The number of users in each group is no more than $\lceil\tfrac{\|\mathcal{U}^{m,g}\|}{G}\rceil$.}

\begin{figure*}[htb]
\setlength{\belowcaptionskip}{-0.5cm}
\setlength{\belowdisplayskip}{-1.2cm}
%\hspace{0.7in}
%\begin{minipage}{0.3\linewidth}
\centering
\subfigure[]{
\label{fig:ic}
\includegraphics[height=0.26\textwidth]{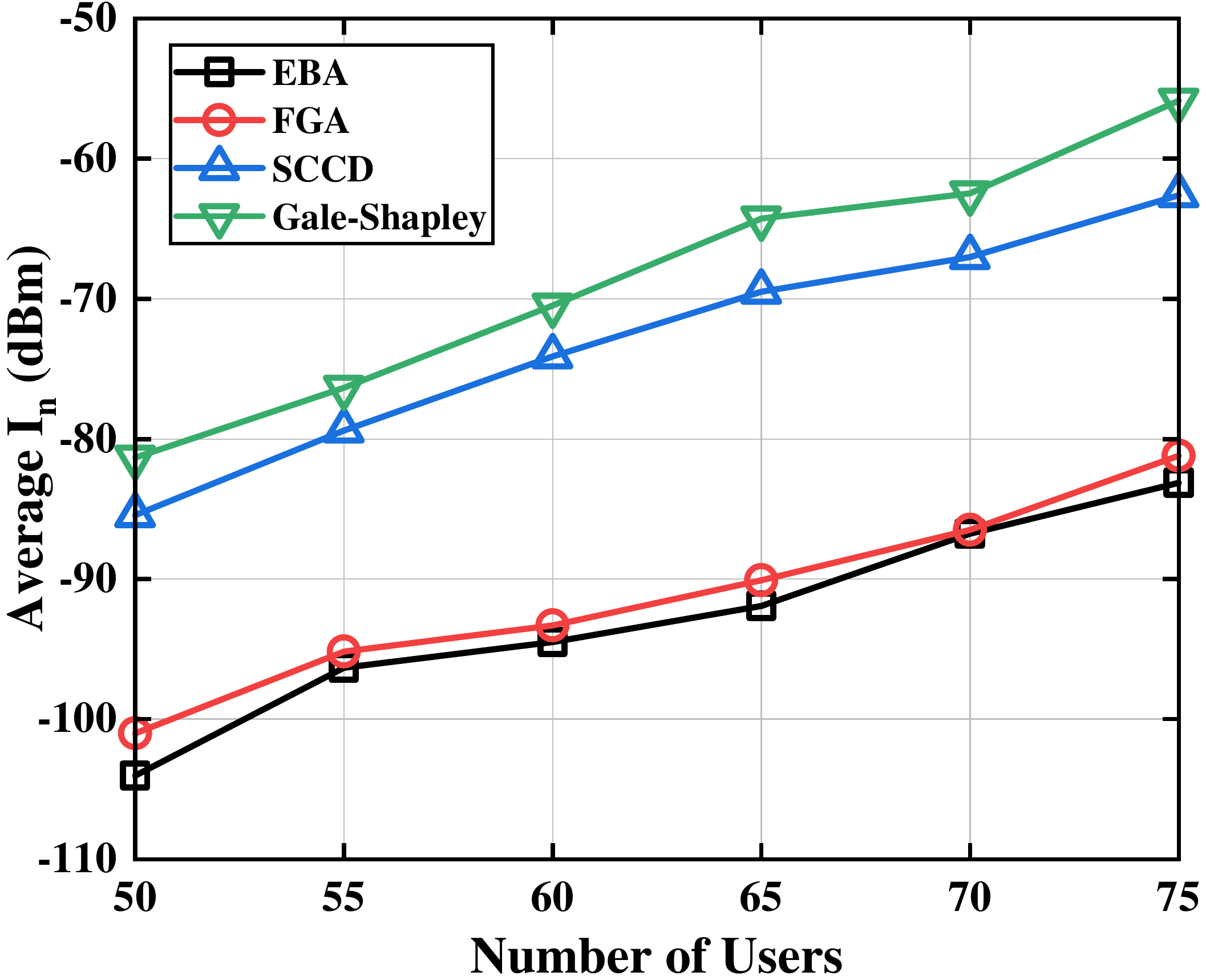}}
\subfigure[]{
\label{figh}
\includegraphics[height=0.26\textwidth]{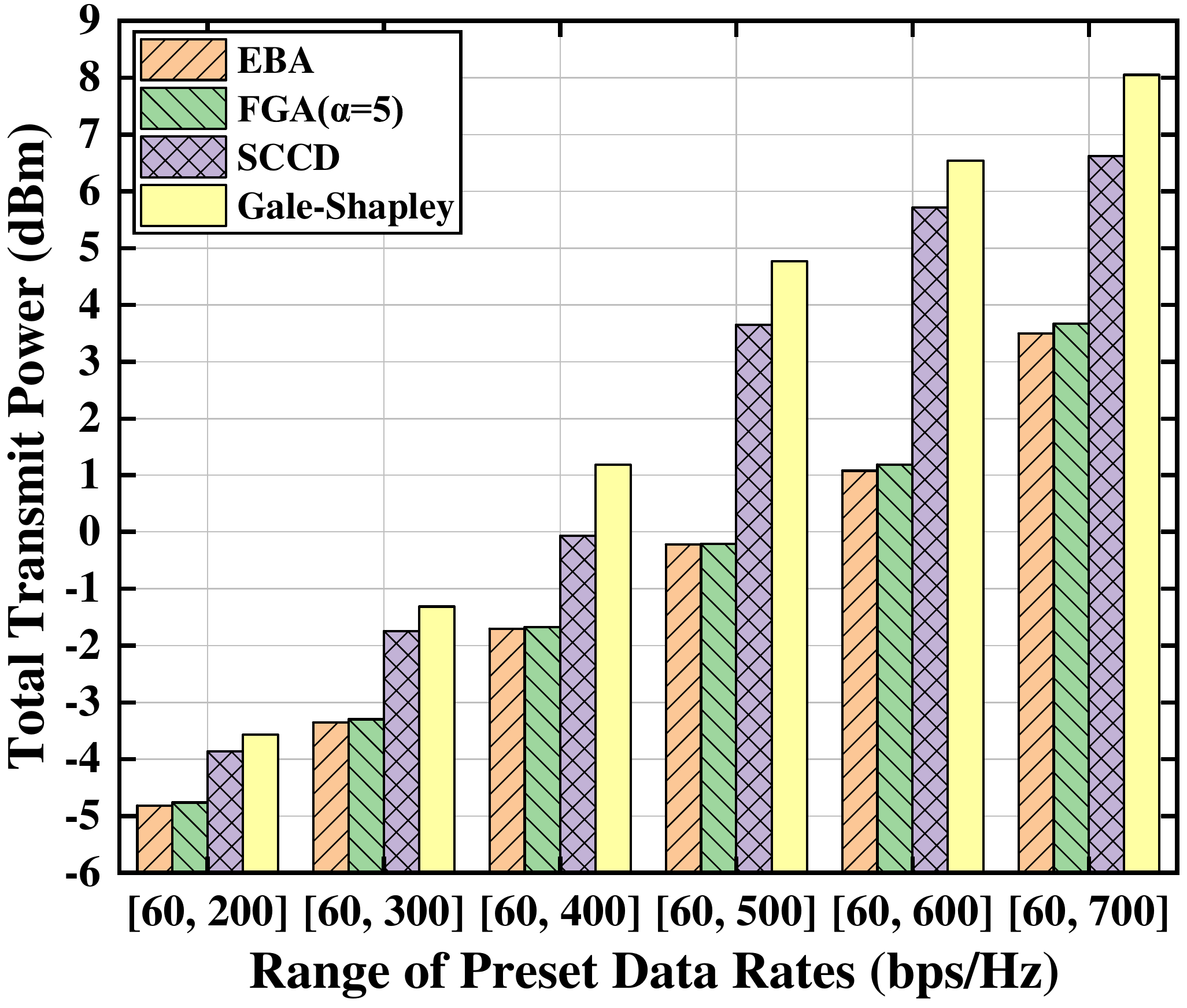}}
%\caption{Detailed Procedure of EBA Strategy(Updated)}
\centering
\caption{Performance Comparison: a. Average  Total Interference vs. Number of Users (10 Groups in Each BS); b. Total Transmit Power vs. Range of Preset Data Rate(15 Groups in Each BS and 150 Users in Total).}
\label{fig:inter-QoS}
%\end{minipage}%
\end{figure*}

As mentioned in Section I,  inter-cell interference is necessary to be considered in multi-cell networks with NOMA. The efficiency of SIC will be seriously affected by inter-cell interference and inter-cell interference is different among different groups in each BS. According to Section II, $I_n$ denotes total interference from the other BSs, and is used to measure inter-cell interference. Fig. \ref{fig:ic} shows average total interference from the other BSs with varied number of users and 10 groups in each BS. As expected, average total interference increases with the number of users for all the four strategies.
Furthermore, the proposed strategies can effectively reduce inter-cell interference compared with the reference strategies. The reason is that, in constraint ($\ref{3619}$c) of our problem ($\ref{3619}$), $I_n$ is an important variable. In other words, in the proposed strategies, not only reducing the transmit power of each BS but also alleviating inter-cell interference {is} taken into consideration. {The interference among users in some group can be partly reduced by selecting users whose channel conditions are more distinctive into the same group (``SCCD'') and avoiding too many users being allocating into the same group(``Gale-Shapley''). However, to reduce inter-cell interference, the cooperation among base stations is important and is not considered in these strategies.}

In Fig. \ref{figh}, we vary the range of preset data rate from [60, 200] to [60, 700] kbps to observe its effect on total transmit power with 15 groups in each BS and 150 users in total, respectively. We can see that the total transmit power increases with the preset data rate for all the four strategies.
{As mentioned in Section I, high target data rates users being assigned into the same group will leads to severe interference among the users in this group. As shown in ($\ref{3613}$b), diverse QoS requirements are taken into consideration of the problem formulation, and are not considered in the reference user grouping strategies, which means that users with high target data rate may be assigned into the same group with the reference user grouping strategies. Then more power is required to satisfied the QoS requirements of those users, which will leads severe interference among the users in these groups.} This result confirms the importance of taking QoS requirements into consideration when grouping users.

\begin{figure*}[htb]
\vspace{-0.6cm}
\setlength{\belowcaptionskip}{-0.5cm}
\setlength{\belowdisplayskip}{-1.2cm}
%\hspace{0.7in}
%\begin{minipage}{0.3\linewidth}
\centering
\subfigure[]{
\label{fignormal:1}
\includegraphics[height=0.26\textwidth]{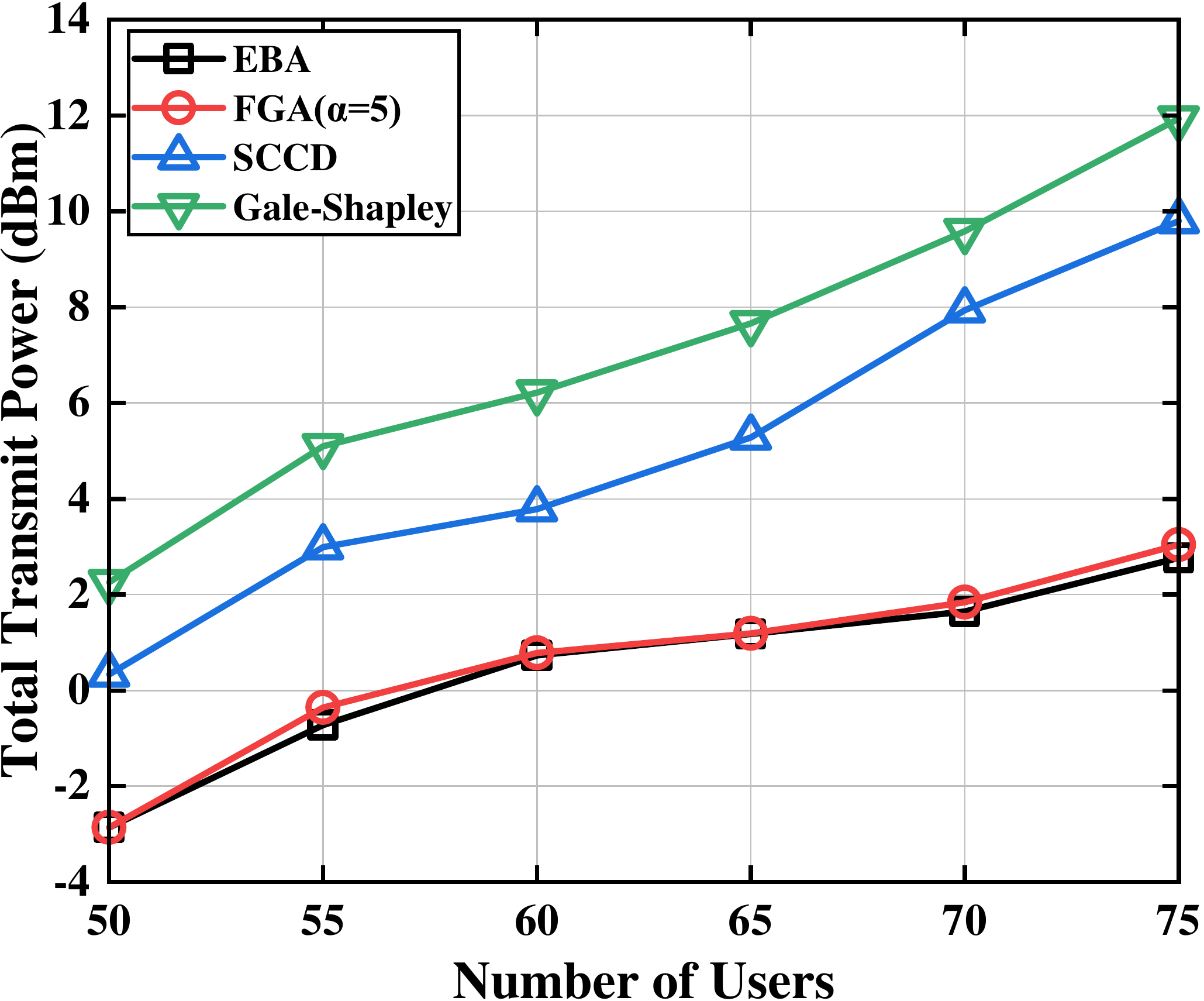}}
\subfigure[]{
\label{fignormal:2}
\includegraphics[height=0.26\textwidth]{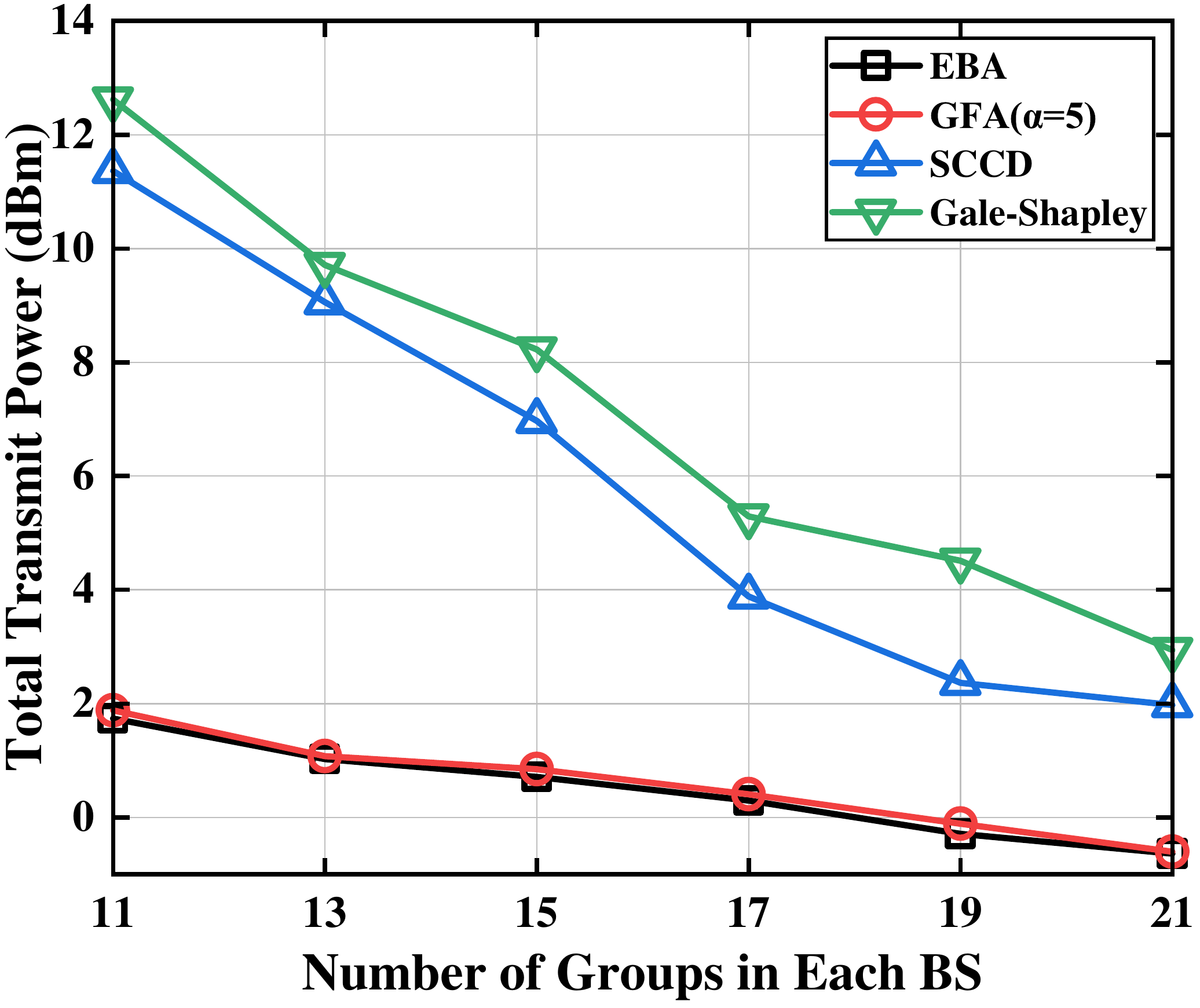}}
%\caption{Detailed Procedure of EBA Strategy(Updated)}
\centering
\caption{Performance Comparison: a. Total Transmit Power vs. Number of Users (10 groups in Each BS); b. Total Transmit Power vs. Number of Groups in Each BS (100 Users in Total).}
\label{fig:usergroup}
%\end{minipage}%
\end{figure*}

Fig. \ref{fignormal:1} shows total transmit power with varied number of users and 10 groups in each BS. Fig. \ref{fignormal:2} shows total transmit power with $G=[11, 13, 15, 17, 19, 21]$ and 100 users in total. As expected, total transmit power increases rapidly as the number of users increases while decreases as the number of groups increases in each BS for all the four curves. The reason is that inter-user interference in each group will rises as the number of users in each group increases as shown in Fig. \ref{fig:ic}. This result confirms that implementing NOMA among all users simultaneously is impracticable. We can also see that total transmit power of the proposed strategies is less than that of the reference strategies. The reason is that the proposed strategies can reduce total transmit power more effectively with the consideration of interference and QoS constraints as shown in problem ($\ref{3619}$).
{As shown in Fig. \ref{fig:ic} and Fig. \ref{figh}, inter-cell interference and diverse QoS requirements {are} affected distinctly by user grouping strategy, and which are not considered in the reference user grouping strategies. As the total number of users increases or the total number of groups decreases, the average number of users in each group increases. The interference among the signals of users served on the same channel will strengthen. The advantage of the proposed user grouping strategies compared with the reference user grouping strategies will be more obvious.}

\begin{figure}[ht]
\vspace{-0.3cm}
\setlength{\abovecaptionskip}{-0.01cm}
\setlength{\belowcaptionskip}{-0.4cm}
	\centering	\includegraphics[width=3in]{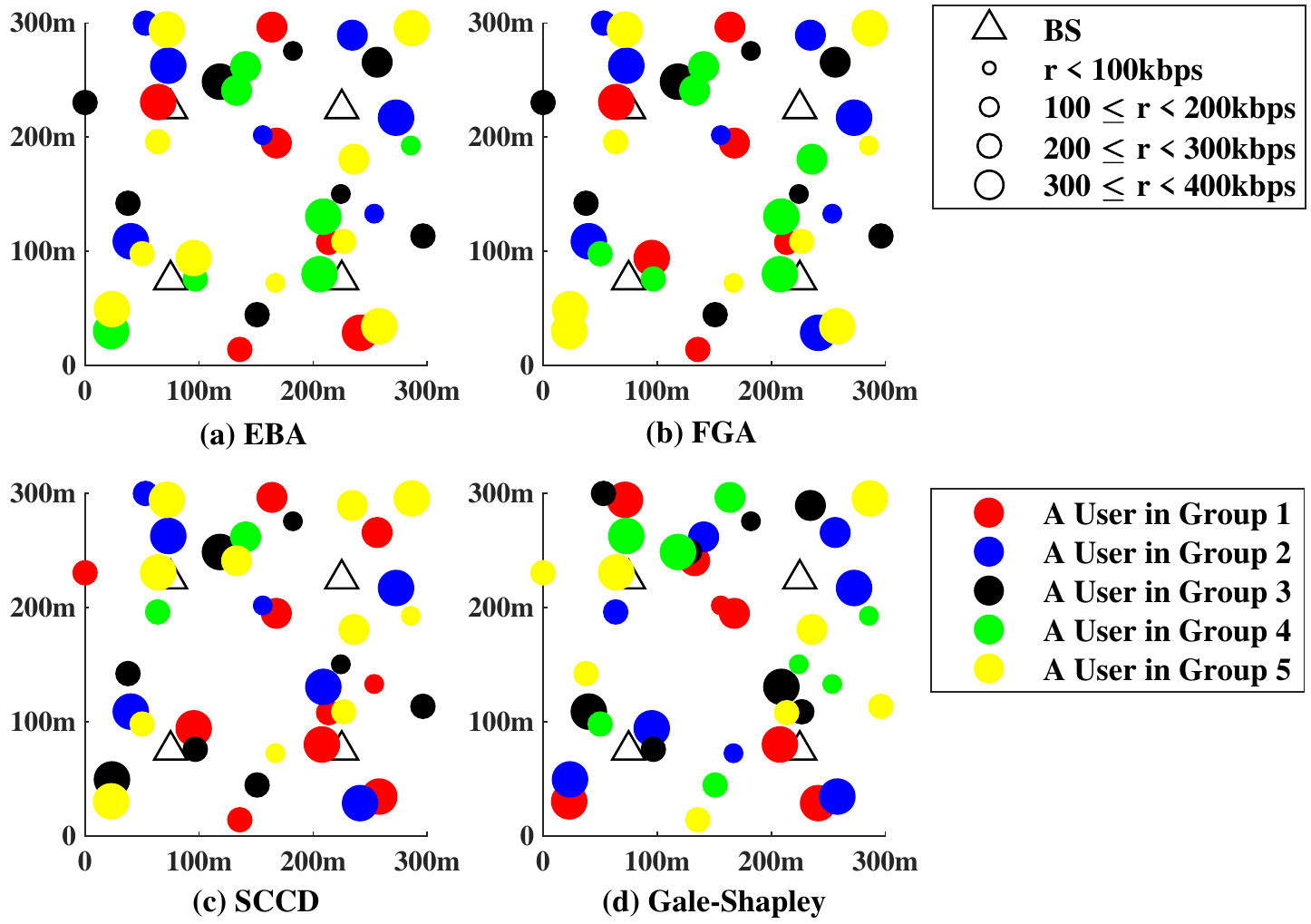}
	\caption{Distribution of Users in 300m $\times$ 300m Rectangular Area(40 Users and 5 Groups)}
	\label{fig:4}
\end{figure}

%Much more power will be consumed if one of these principles is unsatisfied.To show the superiority of the proposed strategies,
According to the analysis in Section I, to reduce inter-user interference, two important principles should be taken into consideration: (a) we should avoid the high target data rates users be assigned into the same group; (b) we should avoid the users with poor channel condition be assigned into the same group. Fig. \ref{fig:4} shows the distribution of BSs and users, where $N=40$ and $G=5$. We narrow the range of user distribution into a 300m$\times$300m rectangular area, where each user is represented by a dot. The users in the same color are in the same group. The distance between the user and the BS indicates this user's channel condition, and the size of each dot indicates this user's target data rate. {As each user prefers the group with less users and each group prefers the user with higher channel gain with Gale-Shapley algorithm, users with high channel gains may be assigned into one group, and the users with poor channel condition being assigned into one group.} In Fig. \ref{fig:4}(d), there are three users close to the edge of the BS range in group $4$ of the higher right BS, which will lead to severe inter-user interference in this group. Although the users with poor channel condition can be avoided to be assigned into the same group with SCCD due to its objective of selecting users whose channel conditions are more distinctive into the same group, the users with high target data rate {are} unavoidable to be assigned into the same group. In Fig. \ref{fig:4}(c), there are three users with high target data rates (more than 200 kbps) in group $5$ of the higher left BS, which will also make inter-user interference in this group severe. By contrast, such inter-user interference is avoided in the proposed strategies as \ref{fig:4}(a) and \ref{fig:4}(b) show. Besides, with much lower computation complexity, the performance of FGA is close to that of EBA.

\section{Conclusion}
In this paper, we study the user grouping problem in downlink multi-cell NOMA systems with QoS constraints based on game theory and graph theory. We prove that the optimal SIC decoding order is the ascending order of CCINR parameter and give the optimal power allocation strategy for any given user grouping strategy. Based on these, we formulate the game model of the problem, and we prove that it can be classified as an exact potential game. We convert the problem for each player, i.e., each BS, to find an appropriate action into the problem of searching for specific negative loops in the graph composed of the users. We can obtain the all-stable solution by searching for those negative loops and updating user grouping strategy. We extend Bellman-Ford algorithm to search for these negative loops. A greedy fast strategy has also been proposed to solve the problem with polynomial time. We show the detailed procedure of reducing total transmit power by the proposed strategy through simulation. Then inter-cell interference can be reduced by the proposed strategies compared with two reference strategies as the simulation results indicate. As inter-cell interference be reduced, the proposed strategies outperform existing grouping strategies on saving power.

In OMA systems, it's well known that a user should associate to the nearest BS. However, this principle might not be exact in multi-cell NOMA systems due to both inter-cell interference and intra-group interference. In practice, a cell-edge user will prefer to associate to the BS where it can be assigned into a group with less interference from the other cochannel cells or users in multi-cell NOMA systems. In the future, we will focus on how to associate cell-edge users to the appropriate BS based on the proposed user grouping strategy in this paper.

\appendices
\section*{Appendix A}

The proof is divided into three steps:

\textbf{Step 1}: {When the SIC decoding order is the ascending order of $S^{m,g}_n$, every user can correctly detect the information of users with {higher} priority detecting order.

According to the NOMA principle, BS $m$ exploits SC and broadcasts signal
$\sum_{n\in\mathcal{U}^{m,g}}\sqrt{p_{n}}s_n$ to users in group $mg$, where $s_n$ denotes the transmitted symbol of user $n$, and $\mathbb{E}[s_n] = 0$ and $\mathbb{E}[| s_n|^2] = 1$ for $\forall n \in \mathcal{U}^{m,g}$.
The signal that user $n$ receives is given by
\begin{equation}\small\label{A_2000}\nonumber
\setlength\abovedisplayskip{1pt}
\setlength\belowdisplayskip{1pt}
y_n\!\!=\!\!H^{m,g}_{n}\!\!\!\!\sum_{i\in\mathcal{U}^{m,g}}\!\!\!\!\sqrt{p_{i}}s_i+\!\!\!\!\!\!\sum_{m'\in\mathcal{M}\setminus {m}}\!\!\!\!\!\!H^{m',g}_{n}\!\!\!\!\sum_{i\in\mathcal{U}^{m',g}}\!\!\!\!\!\!\sqrt{p_{i}}s_i+\mathfrak{n}_{n},~~n\in \mathcal{U}^{m,g},
\end{equation}where $\mathfrak{n}_{n}\thicksim\mathcal{N}(0, \sigma^2)$ is Additive White Gaussian Noise (AWGN). $H^{m,g}_{n}$ denotes the channel coefficient between BS $m$ ($m \in \mathcal{M}$) and user $n $ ($n\in \mathcal{N}$) on subchannel $g $ ($g\in \mathcal{G}$).

With SIC, user $i$ can cancel the interference from the signal of user $n$ if user $i$ can decode any data that user $n$ can decode \cite{You2018Resource}, i.e., $R_{i\dashrightarrow n}\geq R_{n}$. This condition can be written as: for $\forall n, i \in  \mathcal{U}^{m,g}, S^{_{m,g}}_{i}>S^{m,g}_{n}$, we have
\begin{equation}\small
\setlength\abovedisplayskip{1pt}
\setlength\belowdisplayskip{1pt}
\label{re_a602}
\begin{aligned}
&R_{n}=log_2
\Bigg(\!\!1+\tfrac{|H_{n}^{m,g}|^2p_{n}}{|H_{n}^{m,g}|^2\!\!\!\!\!\!\!\!\sum\limits_{_{S^{_{m,g}}_{j}>S^{m,g}_{n}}^{~~j\in\mathcal{U}^{m,g}} }\!\!\!\!\!\!\!\!p_{j}+I_n+\sigma^2}
 \Bigg)\!\!\\
 &\leq \!\!log_2
\Bigg(\!\!1+\tfrac{|H_{i}^{m,g}|^2p_{n}}{|H_{i}^{m,g}|^2\!\!\!\!\!\!\!\!\sum\limits_{_{S^{_{m,g}}_{j}>S^{m,g}_{n}}^{~~j\in\mathcal{U}^{m,g}} }\!\!\!\!\!\!\!\!p_{j}+I_i+\sigma^2}
 \Bigg)=R_{i\dashrightarrow n}.%~~~~~n\in \mathcal{U}^g.
\end{aligned}
\end{equation}It should be noted that the terms of $\sum p_j$ are the same in $R_{n}$ and $R_{i\dashrightarrow n}$. The reason is that, no matter in $R_{n}$ or $R_{i\dashrightarrow n}$, it is always desired to decode the signal of user $n$. When user $i$ decodes the signal of user $n$, all the signals that have not been decoded, i.e., $\!\!\!\!\!\!\!\!\sum\limits_{_{S^{_{m,g}}_{j}>S^{m,g}_{n}}^{~~j\in\mathcal{U}^{m,g}} }\!\!\!\!\!\!\!\!\!\sqrt{p_{j}}s_j$ in $R_{i\dashrightarrow n}$, are regarded as noise, which is the same as this part in $R_{n}$ when user $n$ decodes its own signal.

We have assumed that the SIC decoding condition can be satisfied if every user can decode any data that the user with lower priority detecting order can decode \cite{You2018Resource}, i.e., $R_{i\dashrightarrow n}\geq R_{n}$ and $S^{_{m,g}}_{j}>S^{m,g}_{n}$, for $\forall n,i $ in the same group. This condition can be written as($\ref{re_a602}$), and
($\ref{re_a602}$) is equivalent to
\begin{equation}\small
\setlength\abovedisplayskip{1pt}
\setlength\belowdisplayskip{1pt}
\label{re_a603}
\begin{aligned}
\tfrac{p_{n}}{\!\!\sum\limits_{_{S^{_{m,g}}_{j}>S^{m,g}_{n}}^{~~j\in\mathcal{U}^{m,g}} }\!\!p_{j}+\tfrac{I_n+\sigma^2}{|H_{n}^{m,g}|^2}}\leq
\tfrac{p_{n}}{\!\!\sum\limits_{_{S^{_{m,g}}_{j}>S^{m,g}_{n}}^{~~j\in\mathcal{U}^{m,g}} }\!\!p_{j}+\tfrac{I_i+\sigma^2}{|H_{i}^{m,g}|^2}}.
%~~~~~n\in \mathcal{U}^g.
\end{aligned}
\end{equation}
According to the definition of $S^{_{m,g}}_{n}$, we have $\tfrac{I_n+\sigma^2}{|H_{n}^{m,g}|^2}\geq\tfrac{I_i+\sigma^2}{|H_{i}^{m,g}|^2}$, so ($\ref{re_a603}$) is valid. Therefore, the SIC decoding condition can be met with the proposed SIC decoding order. }

\textbf{Step 2}: Power consumption will be the minimal if we allocate power to users according to ($\ref{a604}$) for any SIC decoding order.
\begin{equation}\small
\setlength\abovedisplayskip{1pt}
\setlength\belowdisplayskip{1pt}
\label{a604}
\begin{aligned}
p^{m,g}_{k}=(2^{r^{m,g}_{k}}-1)(\sum\limits_{_{1\leq i<k}}p^{m,g}_{i}+\tfrac{1}{\varsigma^{m,g}_{k}})
,%~~~~~n\in \mathcal{U}^g.
\end{aligned}
\end{equation}
where all users in group $mg$ have been numbered from $1$ to $\|\mathcal{U}^{m,g}\|$ according to their SIC decoding order, $p^{m,g}_{k}$, $r^{m,g}_{k}$ and $\varsigma^{m,g}_{k}$ stand the transmit power, target data rate and CCINR of the $k$-th users in group $mg$.

{Firstly, it should be noted that according to the principle of SIC, for any SIC decoding order, the SIC decoding order should be the descending order of transmit power. With SIC, the signal of the user with the lowest transmit power in each group would be decoded after the signals of the other users and would not be interfered by the signals of the other users.} Therefore, the minimal power should be allocated to the last user to be decoded in group $ mg $,
i.e. $ p_{1}^{m,g} $ is
{\setlength\abovedisplayskip{1pt}
\setlength\belowdisplayskip{1pt}
\begin{equation}\small
\label{a605}
p^{m,g}_{1}=(2^{r^{m,g}_{1}}-1)\tfrac{1}{\varsigma^{m,g}_{1}}.
\end{equation}}Similarly, the minimal power allocated to the next to last user to be decoded in group $ g $ is
\begin{equation}\label{a606}\small
\setlength\abovedisplayskip{1pt}
\setlength\belowdisplayskip{1pt}
p^{m,g}_{2}\!\!=\!\!(2^{r^{m,g}_{2}}\!-\!1)(\tfrac{1}{\varsigma^{m,g}_{2}}\!+\!p^{m,g}_{1}) .
\end{equation}
By repeating the process, we can get the best power allocation strategy of all users as stated in Theorem \ref{thm1-}.

\textbf{Step 3}: The optimal SIC decoding order is in ascending order of $S^{m,g}_n$, if we allocate power according to step $2$.
We w.l.o.g. assume that the CCINR of the users in group $mg$ have been ordered as $\varsigma^{m,g}_{1}\geq\cdots\geq \varsigma^{m,g}_{\|\mathcal{U}^{m,g}\|}$.

The minimal total transmit power of users in group $\!mg\!$ is
{\setlength\abovedisplayskip{1pt}
	\setlength\belowdisplayskip{1pt}
\begin{equation}\label{a607}\small\nonumber
\begin{aligned}
\setlength\abovedisplayskip{5pt}
	\setlength\belowdisplayskip{5pt}
\!\!\!\!P^{m,g}\!=\!\!\!\!\!\!\!\sum\limits_{^{1\leq k\leq \|\mathcal{U}^{m,g}\|}}\!\!\!\!\!\!\!p_k^{m,g}=\!\!\!\!\sum\limits_{^{1\leq k<
j}}\!\!p_k^{m,g}+p_j^{m,g}+p_{j+1}^{m,g}+\!\!\!\!\!\!\!\!\!\!\sum\limits_{^{j+1< k\leq \|\mathcal{U}^{m,g}\|}}\!\!\!\!\!\!\!\!p_k^{m,g},%\quad g\in \mathcal{G},
\end{aligned}
\end{equation}}where the minimal transmit power of the $j$-th and the $j\!+\!1$-th, $1\leq j<\|\mathcal{U}^{m,g}\|$, in group $mg$ is
\begin{equation}\label{a608}\small\nonumber
\begin{aligned}
\setlength\abovedisplayskip{1pt}
\setlength\belowdisplayskip{1pt}
&p_j^{m,g}\!\!+\!p_{j+\!1}^{m,g}\!
\!=\!\!(2^{r^{m,g}_{j}}\!\!\!\!-\!\!1)(\!\!\sum\limits_{_{1\leq i< j}}\!\!\!p_i^{m,g}\!\!+\!\varsigma^{m,g}_{j})\!\!+\!(2^{r^{m,g}_{j+1}}\!\!-\!\!1)(\!\!\!\!\!\sum\limits_{_{1\leq i<{j+1}}}\!\!\!\!\!p_{i}\!+\!\varsigma^{m,g}_{j+1})\\
&=(2^{r^{m,g}_{j}}\!\!-\!1)(\!\!\!\sum\limits_{_{1\leq i< j}}\!\!p_i^{m,g}\!+\!\varsigma^{m,g}_{j})\!\!+\!(2^{r^{m,g}_{j+1}}-\!1)(\!\!\sum\limits_{_{1\leq i<{j}}}\!\!\!p_{i}+\varsigma^{m,g}_{j+1})\\
&\quad+\!(2^{r^{m,g}_{j+1}}-1)(p^{m,g}_{j+1})\\
%=&\tfrac{2^{r^{_g}_{^j}}\!-\!1}{|H^g_j|^2}\Big(\!\sigma^2\!\!+\!\!\!\!\sum\limits_{^{1\leqi<j}}\!\!\!|H^g_i|^2p^g_i\Big)\!\!+\!\tfrac{2^{r^{_g}_{^{j+1}}}\!-1}{|H^g_{j+\!1}|^2}\Big(\sigma^2\!\!+\!\!\!\!\sum\limits_{^{1\leq i<{j}}}\!\!\!|H^g_i|^2p^g_i\Big)\\
%&+\tfrac{(2^{r^{_g}_{^{j+1}}}\!-\!1)(2^{r^{_g}_{^j}}\!-\!1)}{|H^g_{j+1}|^2}\Big(\sigma^2\!\!+\!\!\!\sum\limits_{^{1\leq i<j}}\!\!|H^g_i|^2p^g_i\Big)
\end{aligned}
\end{equation}

Assume that user $a$ and user $b$ are the $j$-th and the $(j\!+\!1)$-th in group $g$, respectively, and $h_a>h_b$. Then their total minimal transmit power is
\begin{equation}\label{a609}\small\nonumber
\setlength\abovedisplayskip{1pt}
\setlength\belowdisplayskip{1pt}
\begin{aligned}
&p_a\!\!+\!p_b\!=\!(\!2^{r_a}\!\!-\!1)(\!\!\sum\limits_{_{1\leq i< j}}\!\!\!p_i^{m,g}\!\!+\!\!\varsigma_a)\!\!+\!(2^{r_b}\!\!-\!\!1)(\!\sum\limits_{_{1\leq i<{j}}}\!\!p_{i}+\varsigma_b)\!\!+\!\!(2^{r_b}\!\!-\!\!1)p_{a}\\
&=(2^{r_a}-1)(\!\sum\limits_{_{1\leq i< j}}\!\!p_i^{m,g}+\varsigma_a)\!\!+\!(2^{r_b}-1)(\!\sum\limits_{_{1\leq i<{j}}}\!\!p_{i}+\varsigma_b)\\
&\quad+\!(2^{r_b}-1)(2^{r_a}-1)(\!\sum\limits_{_{1\leq i< j}}\!\!p_i^{m,g}+\varsigma_a)\\
%=&\tfrac{2^{r^{_g}_{^j}}\!-\!1}{|H^g_j|^2}\Big(\!\sigma^2\!\!+\!\!\!\!\sum\limits_{^{1\leqi<j}}\!\!\!|H^g_i|^2p^g_i\Big)\!\!+\!\tfrac{2^{r^{_g}_{^{j+1}}}\!-1}{|H^g_{j+\!1}|^2}\Big(\sigma^2\!\!+\!\!\!\!\sum\limits_{^{1\leq i<{j}}}\!\!\!|H^g_i|^2p^g_i\Big)\\
%&+\tfrac{(2^{r^{_g}_{^{j+1}}}\!-\!1)(2^{r^{_g}_{^j}}\!-\!1)}{|H^g_{j+1}|^2}\Big(\sigma^2\!\!+\!\!\!\sum\limits_{^{1\leq i<j}}\!\!|H^g_i|^2p^g_i\Big)
\end{aligned}
\end{equation}

If we exchange the detecting order of user $a$ and user $b$, their total minimal transmit power will be
\begin{equation}\label{a610}\small\nonumber
\setlength\abovedisplayskip{1pt}
\setlength\belowdisplayskip{1pt}
\begin{aligned}
&p_a\!\!+\!p_b\!=\!(2^{r_b}\!\!-\!\!1)(\!\!\sum\limits_{_{1\leq i< j}}\!\!\!p_i^{m,g}\!\!+\!\varsigma_b)\!\!+\!(2^{r_a}\!\!-\!1)(\!\!\sum\limits_{_{1\leq i<{j}}}\!\!\!p_{i}\!+\!\varsigma_a)\!+\!(2^{r_a}\!-\!1)p_{b}\\
&=(2^{r_b}-1)(\!\sum\limits_{_{1\leq i< j}}\!\!p_i^{m,g}+\varsigma_b)\!\!+\!(2^{r_a}-1)(\!\sum\limits_{_{1\leq i<{j}}}\!\!p_{i}+\varsigma_a)\\
&\quad+\!(2^{r_a}-1)(2^{r_b}-1)(\!\sum\limits_{_{1\leq i< j}}\!\!p_i^{m,g}+\varsigma_b)\\
%=&\tfrac{2^{r^{_g}_{^j}}\!-\!1}{|H^g_j|^2}\Big(\!\sigma^2\!\!+\!\!\!\!\sum\limits_{^{1\leqi<j}}\!\!\!|H^g_i|^2p^g_i\Big)\!\!+\!\tfrac{2^{r^{_g}_{^{j+1}}}\!-1}{|H^g_{j+\!1}|^2}\Big(\sigma^2\!\!+\!\!\!\!\sum\limits_{^{1\leq i<{j}}}\!\!\!|H^g_i|^2p^g_i\Big)\\
%&+\tfrac{(2^{r^{_g}_{^{j+1}}}\!-\!1)(2^{r^{_g}_{^j}}\!-\!1)}{|H^g_{j+1}|^2}\Big(\sigma^2\!\!+\!\!\!\sum\limits_{^{1\leq i<j}}\!\!|H^g_i|^2p^g_i\Big)
\end{aligned}
\end{equation}

%user a and user b consume less total transmit power

It is obvious that user $a$ and user $b$ consume less transmit power after exchanging their detecting order. In other word, the $j$-th and the
$(j\!+\!1)$-th in group $g$ will consume less total transmit power if $\varsigma^{m,g}_{j}\!\!<\!\!\varsigma^{m,g}_{j+1}$. Similarly, the channel condition order of all users should satisfy $\varsigma^{m,g}_{k}\!\!<\!\!\varsigma^{m,g}_{k'}$, for $\forall k\!\!<\!\!k', k, k'\!\!\in\!\!\mathcal{U}^{m,g}$. Therefore, if the grouping strategy is settled, the optimal detecting order in each group is also settled. Then if user $\!n$ is assigned into group $g$, the formula for calculating the minimal transmit power can be rewritten as ($\ref{a604}$). Therefore, the proof of \textbf{Step 3} is concluded.

\section*{Appendix B}

According to ($\ref{a605}$) and ($\ref{a606}$), we have
\begin{equation}\small
\setlength\abovedisplayskip{1pt}
\setlength\belowdisplayskip{1pt}
\begin{aligned}
\label{b601}
&\sum\limits_{1\leq k\leq2}\!\!\!p^{m,g}_{k}\!\!=\!p^{m,g}_{1}\!\!+\!\!p^{m,g}_{2}\!=\!(2^{r^{m,g}_{1}}\!\!\!\!\!-\!\!1)\tfrac{1}{\varsigma^{m,g}_{1}}\!\!+\!\!(2^{r^{m,g}_{2}}\!\!\!\!\!-\!\!1)(\tfrac{1}{\varsigma^{m,g}_{2}}\!+\!p^{m,g}_{1})\!\!\!\!\!\!\!\!\!\!\!\!\!\!\!\!\!\!\!\!\!\\
&=(2^{r^{m,g}_{1}}\!\!\!-\!1)\tfrac{1}{\varsigma^{m,g}_{1}}\!+\!\!(2^{r^{m,g}_{2}}\!\!-\!1)\Big(\tfrac{1}{\varsigma^{m,g}_{2}}\!+\!(2^{r^{m,g}_{1}}\!\!\!-\!1)\tfrac{1}{\varsigma^{m,g}_{1}}\!\Big)\\
&=\tfrac{2^{r^{m,g}_{2}}\!-\!1}{\varsigma^{m,g}_{2}}+2^{r^{m,g}_{2}}\tfrac{2^{r^{m,g}_{1}}\!\!\!-1}{\varsigma^{m,g}_{1}}
\end{aligned}
\end{equation}

Combine ($\ref{a604}$) with ($\ref{b601}$), we have
\begin{equation}\small\nonumber
\begin{aligned}
\setlength\abovedisplayskip{1pt}
\setlength\belowdisplayskip{1pt}
\label{b602}
&\sum\limits_{1\leq k\leq3}\!\!\!p^{m,g}_{k}\!\!=\!\!p^{m,g}_{3}\!\!\!+\!\!\!\sum\limits_{1\leq k\leq2}p^{m,g}_{k}\!\!\\
=&\!\!(2^{r^{m,g}_{3}}\!\!\!-\!\!1)(\!\!\sum\limits_{1\leq k\leq2}\!\!\!p^{m,g}_{k}\!\!+\!\!\tfrac{1}{\varsigma^{m,g}_{3}})\!\!+\!\!\tfrac{2^{r^{m,g}_{2}}\!-\!1}{\varsigma^{m,g}_{2}}\!\!+2^{r^{m,g}_{2}}\tfrac{2^{r^{m,g}_{1}}\!\!\!-1}{\varsigma^{m,g}_{1}}\\
=&(2^{r^{m,g}_{1}}\!\!\!-1)\tfrac{1}{\varsigma^{m,g}_{1}}+\!\!(2^{r^{m,g}_{2}}\!-\!1)\Big(\tfrac{1}{\varsigma^{m,g}_{2}}\!+\!(2^{r^{m,g}_{1}}\!\!\!-1)\tfrac{1}{\varsigma^{m,g}_{1}}\Big)\\
=&\tfrac{2^{r^{m,g}_{3}}\!-\!1}{\varsigma^{m,g}_{3}}+2^{r^{m,g}_{3}}\tfrac{2^{r^{m,g}_{2}}\!\!\!-1}{\varsigma^{m,g}_{2}}+2^{r^{m,g}_{2}}2^{r^{m,g}_{3}}\tfrac{2^{r^{m,g}_{1}}\!\!\!-1}{\varsigma^{m,g}_{1}}.
\end{aligned}
\end{equation}
By repeating the process, we can get the minimal transmit power of users in group $mg$:
{%\setlength\abovedisplayskip{1pt}
		\begin{equation}\small\label{b603}
\begin{aligned}
\setlength\abovedisplayskip{1pt}
\setlength\belowdisplayskip{1pt}
&p^{m,g}\!=\!\!\!\sum\limits_{n\in \mathcal{U}^{m,g}}\!\!\!p^{m,g}_{n}\!
=\!\!\!\sum\limits_{n\in \mathcal{U}^{m,g}}\!\!\!\tfrac{2^{r_{n}}\!-\!1}{S^{m,g}_n}\!\!\!\!\!\!\!\prod\limits_{_{S^{_{m,g}}_{i}<S^{m,g}_{n}}^{~~~\!\!i\in\mathcal{U}^{m,g}}}\!\!\!\!\!\!2^{r_{i}}.\\
\end{aligned}
\end{equation}}
Combine ($\ref{2000+}$) and ($\ref{3611}$) with ($\ref{b603}$), we have
{\setlength\abovedisplayskip{1pt}
		\setlength\belowdisplayskip{1pt}
		\begin{equation}\small\label{b604}\nonumber
\begin{aligned}
\setlength\abovedisplayskip{1pt}
\setlength\belowdisplayskip{1pt}
&p^{m,g}\!\!=\!\!\!\!\!\!\sum\limits_{n\in \mathcal{U}^{m,g}}p^{m,g}_{n}\!\!=\!\!\!\!\!\sum\limits_{n\in \mathcal{U}^{m,g}}\!\!\!(2^{r_{n}}\!\!-\!\!1)(\tfrac{I_n+\sigma^2}{|H^{m,g}_{n}|^2})\!\!\!\!\!\!\!\!\!\prod\limits_{_{S^{_{m,g}}_{i}<S^{m,g}_{n}}^{~~i\in\mathcal{U}^{m,g}}}\!\!\!\!\!\!\!\!\!2^{r_{i}}\\
=&\sum\limits_{n\in \mathcal{U}^{m,g}}(2^{r_{n}}-1)(\tfrac{\sigma^2}{|H^{m,g}_{n}|^2})\prod\limits_{_{S^{_{m,g}}_{i}<S^{m,g}_{n}}^{~~~i\in\mathcal{U}^{m,g}}}\!\!2^{r_{i}}\!\!\!\\
&+\!\!\!\!\sum\limits_{n\in \mathcal{U}^{m,g}}\!\!\!\!(2^{r_{n}}\!\!-\!\!1)
(\tfrac{\sum\limits_{m'\in \mathcal{M}\setminus\{m\}}\!\!\!\!\!\!\!\!\!\!\!\!\!|H^{_{m',g}}_{n}|^2p^{m',g}}{|H^{m,g}_{n}|^2})\!\!\!\!\!\!\!\!\prod\limits_{_{S^{_{m,g}}_{i}<S^{m,g}_{n}}^{~~~\!i\in\mathcal{U}^{m,g}}}\!\!\!\!\!\!\!\!\!\!2^{r_{i}}\\
%=&\sum\limits_{n\in \mathcal{U}^{m,g}}(2^{r_{n}}-1)(\tfrac{\sigma^2}{|H^{m,g}_{n}|^2})\prod\limits_{_{S^{_{m,g}}_{i}<S^{m,g}_{n}}^{~~~~~\!i\in\mathcal{U}^{m,g}}}2^{r_{i}}\\
%&+\!\!\!\!\sum\limits_{n\in \mathcal{U}^{m,g}}(\sum\limits_{m'\in \mathcal{M}\setminus
%\{m\}}\!\!\!\!|H^{_{m',g}}_{n}|^2p^{m',g})(\tfrac{2^{r_{n}}-1}{|H^{m,g}_{n}|^2})\!\!\!\!\!\!\!\!\prod\limits_{_{S^{_{m,g}}_{i}<S^{m,g}_{n}}^{~~~~~\!i\in\mathcal{U}^{m,g}}}\!\!\!\!2^{r_{i}}\\
%=&\sum\limits_{n\in \mathcal{U}^{m,g}}(2^{r_{n}}-1)(\tfrac{\sigma^2}{|H^{m,g}_{n}|^2})\prod\limits_{_{S^{_{m,g}}_{i}<S^{m,g}_{n}}^{~~~~~\!i\in\mathcal{U}^{m,g}}}2^{r_{i}}\\
%&+\sum\limits_{m'\in \mathcal{M}\setminus \{m\}}\sum\limits_{n\in
%\mathcal{U}^{m,g}}(|H^{_{m',g}}_{n}|^2p^{m',g})(\tfrac{2^{r_{n}}-1}{|H^{m,g}_{n}|^2})\prod\limits_{_{S^{_{m,g}}_{i}<S^{m,g}_{n}}^{~~~~~\!i\in\mathcal{U}^{m,g}}}2^{r_{i}}\\
=&\sum\limits_{n\in \mathcal{U}^{m,g}}(2^{r_{n}}-1)(\tfrac{\sigma^2}{|H^{m,g}_{n}|^2})\prod\limits_{_{S^{_{m,g}}_{i}<S^{m,g}_{n}}^{~~~\!i\in\mathcal{U}^{m,g}}}2^{r_{i}}\\
&+\!\!\!\!\!\!\!\!\sum\limits_{^{m'\in \mathcal{M}\setminus \{m\}}}\!\!\!\!\!\!\!\!p^{m',g}\!\!\!\!\!\!\sum\limits_{n\in
\mathcal{U}^{m,g}}\!\!\!\!\!(|H^{_{m',g}}_{n}|^2)(\tfrac{2^{r_{n}}-1}{|H^{m,g}_{n}|^2})\!\!\!\!\!\!\!\!\!\prod\limits_{_{S^{_{m,g}}_{i}<S^{m,g}_{n}}^{~~~\!i\in\mathcal{U}^{m,g}}}\!\!\!\!\!\!\!\!\!\!\!2^{r_{i}}.
\end{aligned}
\end{equation}}

The derivation process of ($\ref{3614}$) is finished.

\section*{Appendix C}

To be brief, we define a $k\times k$($1\leq k\leq \|\mathcal{M}\|$)  matrix $\mathbb{X}^{k}$, a $k\times 1$($1\leq k\leq \|\mathcal{M}\|$)  matrix $\mathbb{Y}^{k}$ and a $1\times k$($1\leq k\leq \|\mathcal{M}\|$)  matrix $\mathbb{Z}^{k}$ as follows:
{\begin{equation}\small\label{C001}\nonumber
%\vspace{-0.8cm}
\setlength\abovedisplayskip{0pt}
\setlength\belowdisplayskip{0pt}
\begin{aligned}
\mathbb{X}^k_{ij}=\mathbb{A}^g_{ij}~~~~\forall 1\leq i\leq k, 1\leq j \leq k
\end{aligned}
\end{equation}}
%\vspace{-0.2cm}
{\begin{equation}\small\label{C002}\nonumber
\setlength\abovedisplayskip{0pt}
\setlength\belowdisplayskip{0pt}
\begin{aligned}
\mathbb{Y}^k_{i1}=\mathbb{A}^g_{i(k+1)}~~~~\forall 1\leq i\leq k
\end{aligned}
\end{equation}}
\vspace{-0.2cm}
{\begin{equation}\small\label{C003}\nonumber
\setlength\abovedisplayskip{0pt}
\setlength\belowdisplayskip{0pt}
\begin{aligned}
\mathbb{Z}^k_{1j}=\mathbb{A}^g_{(k+1)j}~~~~\forall 1\leq i\leq k
\end{aligned}
\end{equation}}
%\vspace{-0.2cm}
It is obvious that
{\begin{equation}\small\label{C004}\nonumber
\begin{aligned}
\mathbb{X}^{k+1}=\left[
 \begin{matrix}
   \mathbb{X}^{k} & \mathbb{Y}^k  \\
   \mathbb{Z}^k & -1
  \end{matrix}
  \right]
\end{aligned}
\end{equation}}and{\begin{equation}\small\label{C005}\nonumber
\setlength\abovedisplayskip{1pt}
\setlength\belowdisplayskip{1pt}
\begin{aligned}
\mathbb{X}^{k}=\mathbb{A}^{g}, ~~~~k=\|\mathcal{M}\|
\end{aligned}
\end{equation}}We prove the reversibility of  matrix $\mathbb{A}^g$ using mathematical induction:

(1) $\mathbb{X}^{1}=[-1]$, therefore if $k=1$, matrix $\mathbb{X}^{k}$ is invertible.

(2) If matrix $\mathbb{X}^{k}$ is invertible, the necessary and sufficient condition of the reversibility of  matrix $\mathbb{X}^{k+1}$ is that matrix $-1-\mathbb{Z}^{k}\overline{\mathbb{X}^{k}}\mathbb{Y}^{k}$ is invertible \cite{Wei1998}.

{The channel coefficient $H^{_{m,g}}_{n}$ is a continuous random variable, so according to ($\ref{3615}$), $-1-\mathbb{Z}^{k}\overline{\mathbb{X}^{k}}\mathbb{Y}^{k}$ is also a continuous random variable. Therefore the probability of continuous random variable $\mathbb{Z}^{k}\overline{\mathbb{X}^{k}}\mathbb{Y}^{k}=-1$ is 0. So the probability of continuous random variable $-1-\mathbb{Z}^{k}\overline{\mathbb{X}^{k}}\mathbb{Y}^{k}=0$ is 0. In other words, the probability that matrix $\mathbb{X}^{k+1}$ is irreversible is 0.}

\bibliographystyle{IEEEtran}
\bibliography{bibtex}

%\bibliography{plain}


%
%\begin{IEEEbiography}[{\includegraphics[width=1in,height=1.25in,clip,keepaspectratio]{qqdx.pdf}}]{Fengqian Guo}
%\end{IEEEbiography}}

\end{document}